\documentclass[a4paper,fleqn]{cas-sc}

\usepackage[authoryear]{natbib}
\usepackage{float}
\usepackage{soul}
\newcommand{\karman}{K{\'{a}}rm{\'{a}}n}
\newcommand{\faspace}{\ensuremath{(f^*, A^*)}}

\usepackage[final]{changes}
\definechangesauthor[name={Reviewer \#1}, color=blue]{R1}
\definechangesauthor[name={Reviewer \#2}, color=cyan]{R2}
\definechangesauthor[name={Reviewer \#3}, color=orange]{R3}
\definechangesauthor[name={Authors}, color=red]{A}

\newcommand{\fst}{f_{\text{St}}}

\newcommand{\diskthickness}{\ensuremath{\delta_{\text{d}}}}

\newcommand{\by}{\ensuremath{\times}\,}

\begin{document}
\let\WriteBookmarks\relax
\def\floatpagepagefraction{1}
\def\textpagefraction{.001}
\shorttitle{Wake of oscillating cylinder at low Re}
\shortauthors{Yang, Masroor and Stremler}

\title [mode = title]{The wake of a transversely oscillating circular cylinder in a flowing soap film at low Reynolds number}

\author[1]{Wenchao Yang}[
    orcid=0000-0002-9963-8682]
\ead{wenchao.yang@queensu.ca}
\author[2]{Emad Masroor}[
    orcid=0000-0002-8770-1429]
\ead{emad@vt.edu}
\author[3]{Mark A. Stremler}[
    orcid=0000-0001-5785-7732]
\ead{stremler@vt.edu}
\cormark[1]
\address[1]{Department of Mechanical and Materials Engineering, Queen’s University, Kingston, Ontario K7L 3N6, Canada}
\address[2]{Engineering Mechanics Program, Virginia Tech, Blacksburg VA 24061, USA}
\address[3]{Department of Biomedical Engineering and Mechanics, Virginia Tech, Blacksburg, VA 24061, USA}
\cortext[cor1]{Corresponding author.}

\begin{abstract}
An inclined gravity-driven soap film channel was used to study the wake patterns formed behind a transversely oscillating cylinder at $Re =235 \pm 14$.  The natural frequency of vortex shedding from a stationary cylinder, $\fst$, was used to identify the oscillation frequencies of interest.
The (dimensionless) frequency, $f^*=f/\fst$, and amplitude, $A^*=A/D$, of the cylinder's motion was varied over a large portion of the fundamental synchronization region (i.e., for $f^* \approx 1$), and a `map' of wake patterns was constructed in $\faspace$ space. Lock-on between the frequency of the cylinder's motion and the dominant frequency of the resulting vortex wake was observed for a large range of this parameter space, predominantly manifested as synchronized `2S' and `2P' wake modes. 
Synchronized `P+S', `2T', and `transitional' wakes were also found in smaller regions of parameter space.  
Unsynchronized `coalescing' and `perturbed von~\karman' wakes were observed as the oscillation frequency became sufficiently different from $\fst$.  
The wake patterns and vortex formation processes found in this study, 
particularly \deleted[id=A]{for} the 2P mode wakes, 
\added[id=R1]{bear a strong resemblance to previous three-dimensional experimental results, }
\deleted[id=R1]{are more similar to those observed by \cite{Williamson1988} in three-dimensional experiments than those found by \cite{Leontini2006} in two-dimensional simulations,}
despite the physical constraint from the soap film that 
\replaced[id=R2]{limits}{ prevents} three-dimensional effects in the wake.
\end{abstract}

\begin{graphicalabstract}
\includegraphics{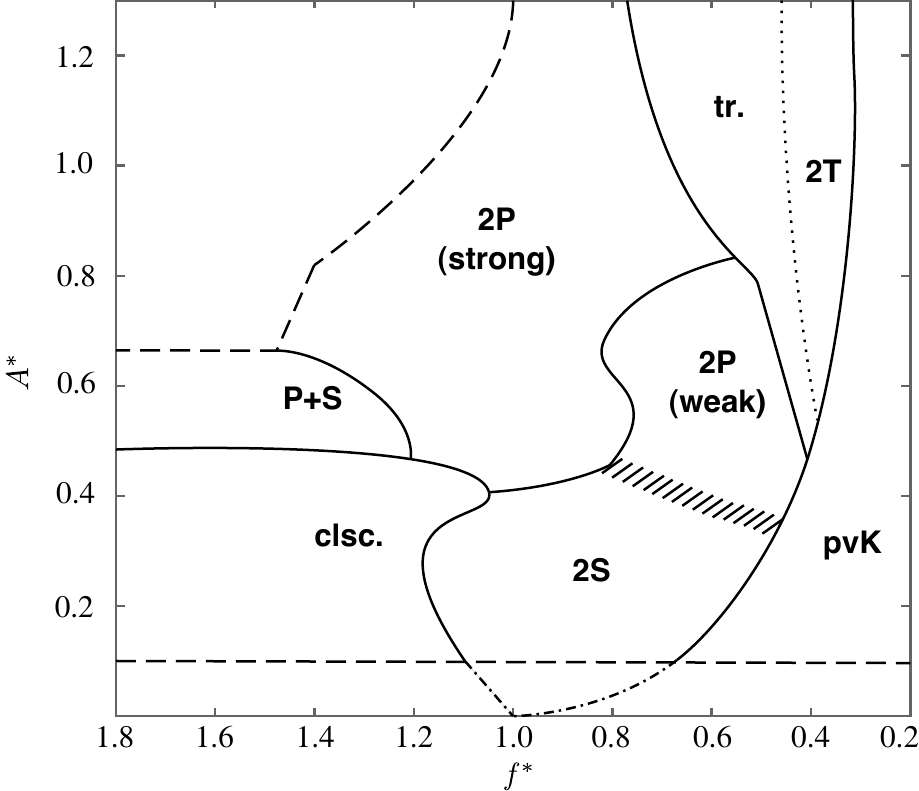}
\end{graphicalabstract}


\maketitle
\section{Introduction}
\begin{figure}
    \centering
    \includegraphics[scale=1]{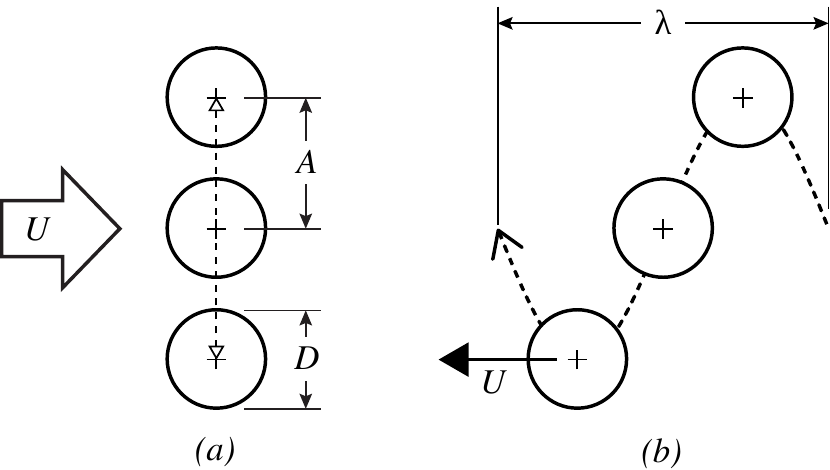}
    \caption{Circular cylinder oscillating (a) transversely relative to a background flow or (b) along a sinusoidal path through a quiescent fluid. }
    \label{fig:cylinder}
\end{figure}

Steady, uniform flow moving with speed $U$ past a fixed circular cylinder with diameter $D$ produces, for a wide range of Reynolds number\added[id=A]{s}, an unsteady wake consisting of coherent vortical structures that are shed from the cylinder with a clearly defined frequency, $\fst$; see, e.g., the review by \citet{Williamson1996} and the textbook by \cite{Zdravkovich1997}.  
When the cylinder is also allowed to oscillate transverse to the flow direction in response to the periodic vortex shedding --- or, as is the case in the present work, when it is forced to oscillate at a prescribed frequency, $f$, and amplitude, $A$ --- the resulting wake structure can be significantly more complicated than when the cylinder is held stationary \citep{Williamson1988}. Equivalent flow behavior is observed when the cylinder is pulled through a fluid at rest along a sinusoidal path with amplitude $A$ and wavelength $\lambda = U/f$, where $U$ is the constant cylinder speed in the direction of the path's centerline, as illustrated in Fig.~\ref{fig:cylinder}; this approach is, in fact, the one used by \citet{Williamson1988}. 
In both cases, the velocity of the cylinder with respect to the background flow is given by
\begin{equation}
	(u,v) = \left(-U, 2\pi f A \cos(2\pi f t) \right),
\end{equation}
where $u$ and $v$ are the streamwise and transverse velocity components, respectively.  
This classic problem is relevant to a wide range of applications, including oil riser motion \citep{Wu2012}, aeroelastic stability \citep{Zhang2017}, wind turbines \citep{Heinz2016}, and small-scale energy harvesting \citep{Antoine2016}, among others. 

The oscillating cylinder problem can be characterized by the following three dimensionless groups: the Reynolds number, $\text{Re} = U D / \nu$, where $\nu$ is the kinematic viscosity of the fluid; the normalized oscillation amplitude $A^* = A/ D$; and the frequency ratio $f^*=f/\fst$. 
The natural frequency of vortex shedding from a single stationary cylinder, $\fst$, is often expressed in terms of the Strouhal number, $\text{St} = \fst D/U$, which is an empirically known function of the Reynolds number \cite[see, e.g.,][]{Williamson1998}.
In some studies, the wavelength ratio, $\lambda^* = \lambda/D$, or, equivalently, the reduced velocity, $U^* = U/(f D)= \lambda/D$, is used instead of $f^*$. Note that $f^* = \text{St}/U^*$, so that for the wide range of Re over which St is approximately constant the parameters $U^*$ and $1/f^*$ can be used interchangeably; however, for $\text{Re}\lesssim 1000$, as considered here, the relationship between these parameters varies with the Reynolds number.   

Two different methodologies have been used to investigate the flow-induced vibration of a circular cylinder submerged in a steady, uniform fluid flow. In the \textit{free vibrations} approach, one considers the response of an elastically-mounted cylinder to the fluid forcing; see \cite{Williamson2004} for a comprehensive review of these `Vortex-Induced Vibrations' (VIV). The independent variable in these experiments is usually the reduced velocity $U^*$, which relates the time scale of the cylinder's oscillation, $\tau = f^{-1}$, to the convective time scale, $D/U$. A key result of such studies is the maximum amplitude of motion experienced by the cylinder; in other words, free vibration studies often aim to determine the maximum amplitude of flow-induced vibration for a given system and the value(s) of the reduced velocity at which this maximum occurs.  These parameters are of considerable practical importance.

The second approach to studying flow-induced vibrations, which was the one adopted for this study, is the \textit{controlled oscillations} approach \citep{Bishop1964, Williamson1988}. In this approach, the cylinder is prescribed to move at particular values of frequency, $f$, and amplitude, $A$, for a specified inflow velocity, $U$. Studies that use this approach are able to access as much of the frequency-amplitude parameter space as experimental (or computational) considerations allow, rather than being constrained by the frequency and amplitude response of a given system, which is typically not known in advance. By judiciously choosing the region of frequency-amplitude space to explore (i.e., frequencies that one wishes to suppress or hopes to excite), these studies serve as a laboratory for artificially-constructed `flow-induced' vibrations. If the amplitude and frequency of a cylinder undergoing controlled oscillation are made to mimic the response of the same cylinder under free vibrations (at a certain reduced velocity), the two problems are  commensurate, and one can provide insight into the other \citep{Govardhan2000,Jiao2018}. Thus, controlled oscillations allow one to create `what-if' scenarios of practical importance that may be difficult to test in the laboratory using free-vibration experiments.

\begin{figure}
	\centering
		\includegraphics[width=165mm]{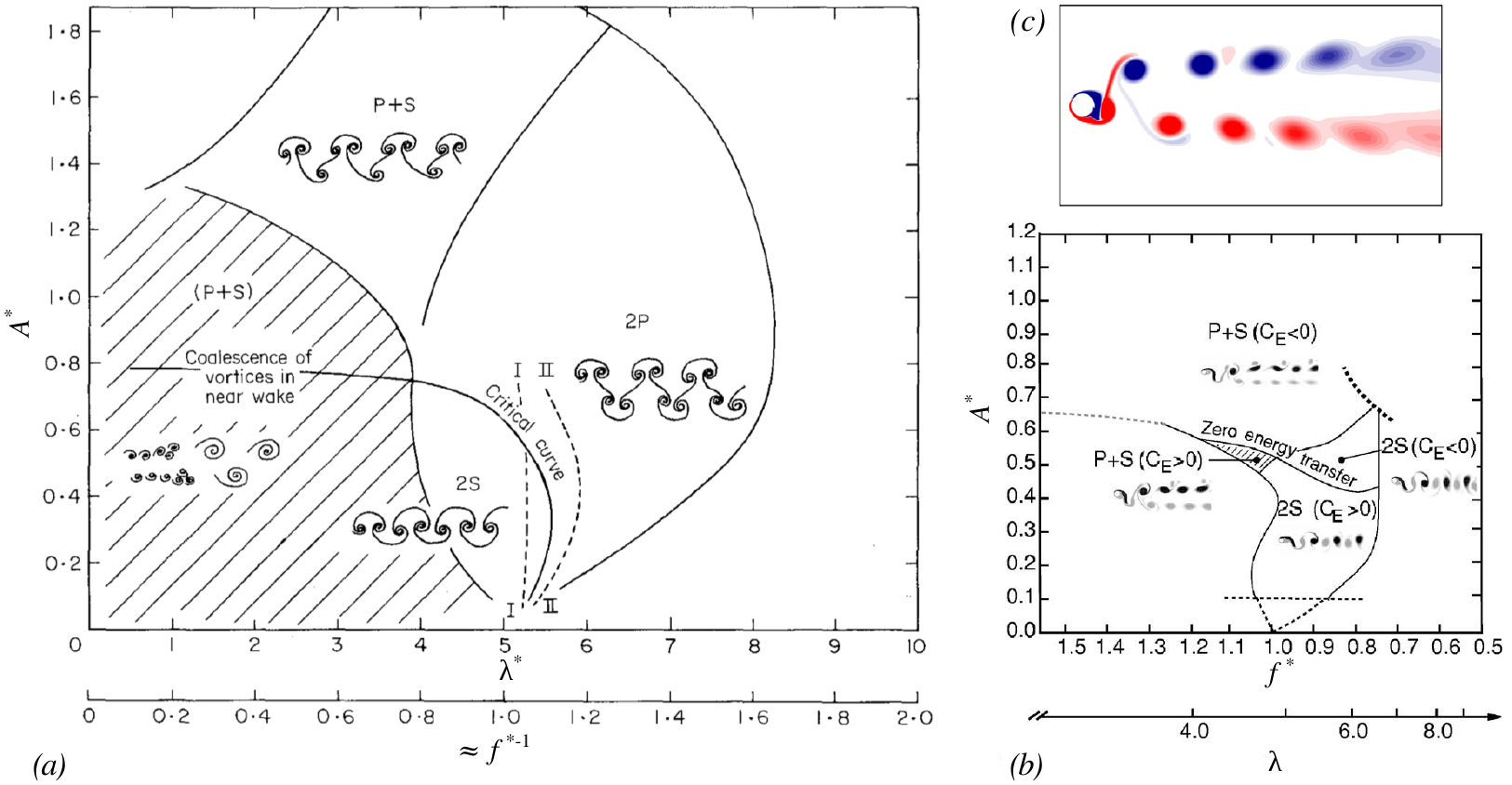}
	\caption{Non-dimensionalized frequency-amplitude `map' of wake patterns for a transversely oscillating cylinder from previous studies: (a) patterns for a cylinder towed sinusoidally in a water tank at $300 < Re < 1000$, with the wake visualized using aluminum particles on the surface, from \cite{Williamson1988}, and  (b) patterns from a two-dimensional computational simulation at $Re = 300$ using a body-fitted grid with an external force term to account for cylinder's transverse acceleration, from \cite{Leontini2006}.
\added[id=R2]{(c)~A weak 2P mode wake in two-dimensional flow at $\text{Re}=300$, from figure~21(c) in \cite{Dorogi2020}.}}
	\label{fig:WR-L}
\end{figure}

Building on the seminal work of \citet{Bishop1964}, \citet{Williamson1988} conducted an extensive, systematic study of flow past a circular cylinder undergoing controlled oscillations over a large range of $f^*$ and $A^*$ values for $300 < Re < 1000$. 
They found that different exotic vortex patterns appear in the cylinder wake for distinct regions of the $\faspace$ parameter space.
In the fundamental synchronization region (i.e., for $f^* \approx 1$), three main patterns were identified: 2S wakes, with two (single) vortices shed per cycle of oscillation; P+S wakes, with a pair of vortices and a single vortex shed per cycle; and 2P wakes, with two pairs of vortices shed per cycle of oscillation. 
This general organization of the $\faspace$ parameter space was confirmed with a highly-resolved study by \cite{Morse2009} at $\text{Re}=4000$, who identified an additional `2P overlap' regime containing wakes with two pairs of very different strength vortices \added[id=A]{(a `weak 2P' mode)}  that appeared intermittently with the typical 2P wake.
\cite{Leontini2006} conducted a similar two-dimensional computational study for $\text{Re} = \{ 100,200,300 \}$ and found the primary synchronization region to be populated only by 2S and P+S wakes. 
The results of the two studies by \cite{Williamson1988} and \cite{Leontini2006} are summarized in figure \ref{fig:WR-L}\added[id=R1]{(a,b)}.
\deleted[id=R2]{The absence of 2P wakes at low values of Re in the computational study has been attributed to the inherent two-dimensionality of the flow. } 

\added[id=R2]{One noticeable difference between the results in figure \ref{fig:WR-L}(a,b) is the presence of  2P mode wakes in the three-dimensional experiments at higher Re values.  
By  forcing the flow to be planar, \cite{Leontini2006} appeared to completely suppress the occurrence of 2P wakes.  \cite{Williamson1988} also noted that in their experiments the  symmetry-breaking P+S mode was observed to replace the symmetric 2P mode  for $\text{Re}<300$.   
It is generally accepted that the 2P mode wake is absent in low Reynolds number and/or two-dimensional wakes (see, e.g., low-Re experiments by \cite{Griffin1974} and computations by \cite{Blackburn1999}).  
However, \cite{Placzek2009} have described the appearance of a weak 2P mode for forced oscillations in a 2D computational model at $\text{Re}=100$, and   \cite{Dorogi2020} show a weak 2P mode occurring in a 2D computational model at $\text{Re}=300$ under both forced and free oscillations.  An illustration of the structure observed by \cite{Dorogi2020} is shown in figure \ref{fig:WR-L}(c).
}

The forces experienced by isolated bodies undergoing flow-induced vibrations are largely caused by vortex shedding at the rear of the body and the subsequent organization of the shed vorticity in the wake. Thus, changes observed in the forces experienced by an immersed body have been attributed to changes in the presence, nature, intermittency, and magnitude of vortex shedding. Abrupt changes in the magnitude of oscillations experienced by a freely-vibrating cylinder while a controlling parameter is varied were first documented by \cite{Bishop1964}. \citet{Williamson1988} provided preliminary evidence that this `jump' in forcing may be related to a change from the 2S to the 2P mode of vortex shedding, a relationship that has been confirmed in free-vibration experiments with a flexibly-mounted circular cylinder in a wind tunnel 
\citep{Brika1993} and in controlled-oscillation experiments with a rigid cylinder in a water tunnel \citep{Morse2009}.
Thus, investigations of wake patterns\added[id=A]{, particularly regarding the appearance (or absence) of the 2P mode,} are relevant for understanding and predicting the fluid forces on oscillating bluff bodies.  

In this work, we extend previous studies by considering the wake patterns formed behind an oscillating cylinder in an experimental system at $\text{Re}\approx 235$ for the parameter ranges $0.2 < f^* < 1.8$ and $0.1 < A^* < 1.3$. We restrict the flow to be (quasi) two-dimensional by conducting the experiments in a gravity-driven soap film. Our approach can be considered a complement to the 3D experimental work of \citet{Williamson1988} that facilitates comparison with the 2D computational study of \citet{Leontini2006}.  
The region of parameter space examined here overlaps much of that explored by \cite{Williamson1988} and includes the entirety of the region explored by \cite{Leontini2006}\added[id=A]{, as shown in figure \ref{fig:unlabeledmap}}. 

Soap films have been used as a `laboratory' for studying two-dimensional flows since their introduction by the seminal work of Couder \citep{Couder1984,Couder1986,Couder1989}.  The utility of soap film systems was enhanced by the introduction of the flowing film `tunnel' by \citet{Gharib1989}. Vertical, gravity-driven soap films were first utilized by \cite{kellay1995prl}, and \citet{Georgiev2002} extended the available parameter space by designing an inclined flowing film system. 
\cite{Auliel2015} have shown that the equations describing the motion of a flowing soap film are analogous to the two-dimensional incompressible Navier--Stokes equations, provided that the Marangoni effect due to surface tension is small. 
A number of important hydrodynamics results have been shown or replicated in soap films, including those by \cite{Yang2001}, who showed that the streamlines produced by steady flow over a backward-facing step correlate well with numerical simulations; \cite{Vorobieff1999}, who found that the Strouhal-Reynolds number relationship for a stationary circular cylinder in a soap film corresponds well to the two-dimensional St-Re relationship for a circular cylinder in a viscous fluid; \cite{Wu2004}, who determined that the time-varying angle of separation from a stationary circular cylinder agreed with results from a spectral numerical method; and \cite{Yang2019}, who established experimental values of critical spacing between tandem cylinders and showed good agreement with two-dimensional simulations. 

The remainder of the manuscript is organized as follows.  In Section \ref{sec:methods} we describe the soap film system and the cylinder oscillation mechanism, explain the flow visualization and velocity measurement techniques, and detail the method used to estimate the Reynolds number.  Section \ref{sec:results} documents the results of our exploration of the parameter space, arranged according to the type of wake pattern observed.  In this soap film system and over the range of parameters covered by this study, documented wake patterns include `2S', weak and strong `2P', `P+S', `2T', `transitional', `coalescing', and `perturbed von~\karman' types.  
Finally, section \ref{sec:conclusions} sets out the main physical insights and conclusions that can be drawn from this study regarding low-Reynolds number flows past transversely-oscillating circular cylinders.

\section{Methods}
\label{sec:methods}

The results presented in section~\ref{sec:results} were obtained from experiments conducted in a flowing soap film channel that was developed following the design introduced by \cite{Georgiev2002} and informed by the work of \cite{jia2008prl} and \cite{Wang2010}.

\subsection{Soap film apparatus}
\label{subsec:soapfilm}

\begin{figure}
	\centering
  \includegraphics[scale=1]{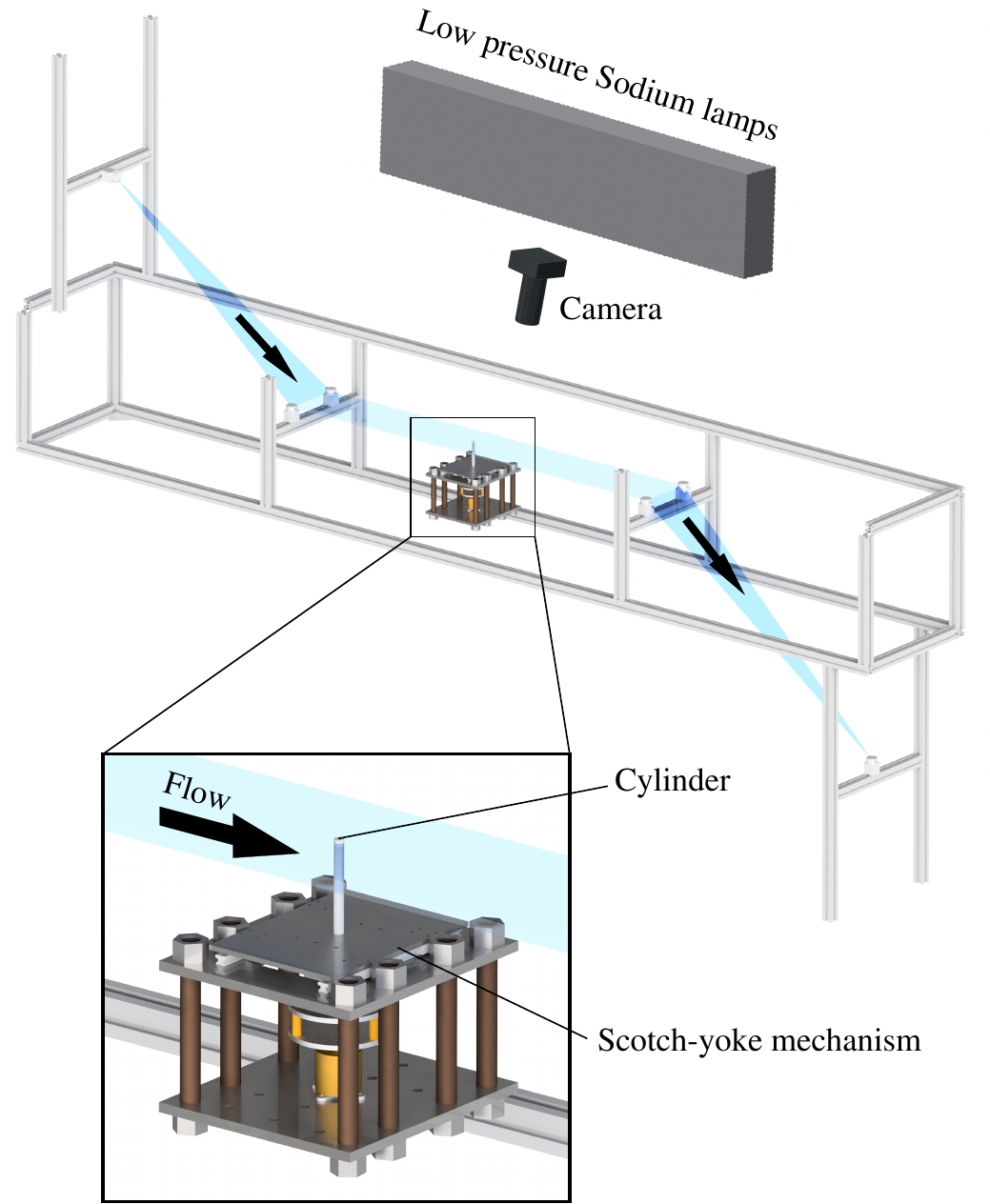}
  \caption{A schematic of the experimental apparatus used for this study. The translucent blue area represents the (approximately) 2-dimensional region occupied by the soap film, with the black arrows representing its gravity-driven flow. Soap solution flows between a valve at the top of the frame and a bucket at the bottom (not shown). The sodium lamp is positioned such that the interference fringe pattern reflected from the upper and lower surfaces of the soap film is captured by the high-speed camera positioned between the lamp and the soap film. A scotch-yoke mechanism, to which an acrylic cylinder is attached, is mounted inside the frame, but isolated from it using a vibration isolation table (not shown). The free end of the cylinder protrudes through the soap film when the system is in operation.}
\label{fig:schematic}
\end{figure}

Figure~\ref{fig:schematic} shows a schematic of our inclined gravity-driven soap film system, which is identical to that used by \cite{Yang2019}. To generate a flowing film, soap solution consisting of 1\% (by volume) Dawn Escapes 
dish soap (Proctor \& Gamble) mixed in water was first pumped to a top reservoir, consisting of an open plastic container held inside a larger, open, overflow container; 
the solution was continually pumped from the overflow into the main container to maintain a constant head in the reservoir throughout each experiment. A valve at the outlet of the reservoir provided control of the volumetric flow rate. A soap film was developed between two nylon guide wires (with diameter 1~mm) that were tensioned at the bottom of the system by a spring. The guide wires were separated using four horizontal nylon pull lines (with diameter 0.3 mm) to form a 58 cm long triangular expansion section, a 10~cm wide by 1~m long rectangular test section, and a 56~cm long triangular contraction section. The expansion section was inclined 57$^{\circ}$ from the horizontal axis, the test section had a 14$^{\circ}$ inclination angle, and the contraction section had a 68$^{\circ}$ inclination angle. The outline of the blue region in figure \ref{fig:schematic} shows the overall shape of the `channel' to which the soap film was thus confined.  We were able to vary the flow speed in the inclined soap film over the range $0.5\,\text{m/s} \lesssim U \lesssim 1.2\,\text{m/s}$. 
The inclination angle of 14$^{\circ}$ in the test section was the minimum angle for which out-of-plane displacement of the film (by gravity) was fully obscured by the nylon guide wires when viewed from the side. In this configuration the curvature of the film surface was approximately 0.4 m$^{-1}$, which we consider to be negligible.

An acrylic rod with diameter $D=6.35$\,mm was mounted on a plate oriented parallel to the test section of the soap film channel, so that the cylinder penetrated the soap film at a 90$^{\circ}$ angle.  The cylinder was placed approximately 30\,cm from the entrance of the rectangular test section.   Wakes were imaged as described in section~\ref{subsec:viz}.  For the oscillation experiments, cylinder motion was driven by a Scotch-yoke mechanism (see section~\ref{subsec:oscillation}).

The validity of using this soap film flow as an analog for the two-dimensional flow of a Newtonian fluid requires that the elastic Mach number, $M_e \equiv U/U_M = O(10^{-1})$ \citep{Auliel2015}, where $U_M$ is the Marangoni wave speed given by $$U_M \equiv \sqrt{2 E_M/(\rho h)}.$$
The fluid density is $\rho$, $h$ is the \added[id=A]{(average)} thickness of the soap film, and $E_M$ is the Marangoni elasticity.
The average flow speed for the experiments reported here was determined to be $U\approx 0.68$\,m/s (see Section~\ref{subsec:flowspeed}). By mass conservation, this average flow speed corresponds to an average film thickness of $h \approx 3\,\mu$m.  
\cite{Kim2017} have shown that for flowing soap film systems similar to the one used here, the Marangoni elasticity is $E_M \approx 22$\,mN/m, and we have assumed this value in our estimate of $U_M$. Taking 
$\rho \approx 1$\,g/cm$^3$, the Marangoni wave speed for these experiments is  estimated to be $U_M \approx 3.8$\,m/s, which gives an elastic Mach number value of $M_e  \approx 0.2$.  We thus have $M_e = O(10^{-1})$ for the experiments reported here, supporting the interpretation of these results as corresponding to those for a two-dimensional Newtonian flow.


\subsection{Flow visualization and quantification}
\label{subsec:viz}
\begin{figure}
	\centering
  \includegraphics[scale=.9]{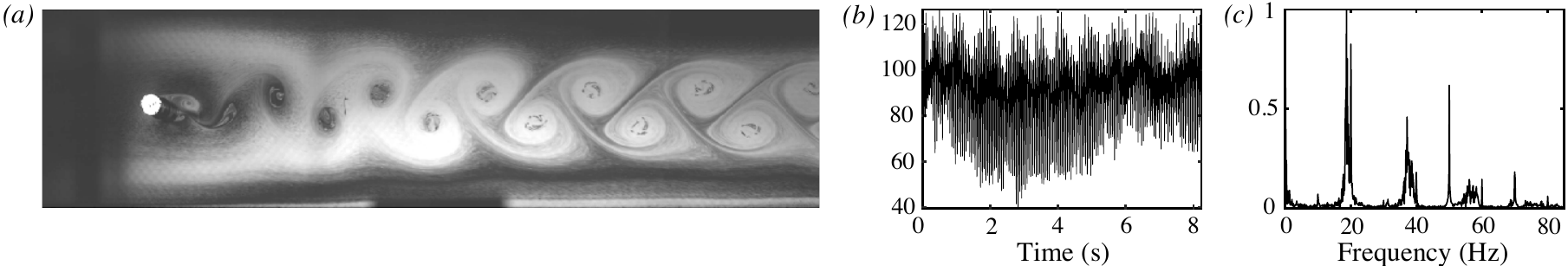}
  \caption{(a) Flow visualization for the wake of a stationary circular cylinder (image size: 2352 \by 600 pixels $\approx 39.2D \by 10D$). A thin white rectangle (1 \by 49 pixels) marks the region used in the quantitative analysis.  (b) Time trace of the (unfiltered) mean intensity of the rectangular region. 
  (c) Power spectrum of the mean intensity, showing a primary peak at 18.67\,Hz.}
\label{fig:wakeFFT}
\end{figure}

The wake patterns were documented by recording videos 8.2 seconds long at 170 frames per second for a total of 1400 frames per case, with each frame having an image resolution of 2352 $\times$ 600 pixels. During each recording, the flow speed $U$, cylinder oscillation frequency $f$, and cylinder oscillation amplitude $A$ were held fixed.  
A monochromatic light source positioned perpendicular to the experimental plane was used to illuminate the soap film, and the reflected patterns of light were recorded with a high-speed camera positioned near the light source. Three low-pressure sodium lamps (SOX 180, Philips; wavelength 590 nm), which were wired to each phase of an alternating current power source, provided monochromatic illumination of the soap film. The phase delay between the lamps produced a nearly constant light intensity; the small fluctuations in intensity had a frequency of 120Hz, as documented in \cite{Yang2019}.  The camera (Falcon 4M60, Teledyne DALSA) recorded the interference fringe patterns produced by the light source as it reflected off the top and the bottom surfaces of the soap film; the resultant changes in the scalar field of light intensity were caused by variations in the thickness field of the soap film. \added[id=R2]{Figure \ref{fig:wakeFFT}(a) shows a sample of the flow visualizations obtained in our experiments.} 

\added[id=R2]{The thickness variations enabling the flow visualization are produced by the flow physics, but these variations are not expected to influence the interpretation of the flow as being two-dimensional and incompressible.   \cite{Wu2001} and \cite{Georgiev2002} have demonstrated that a  flowing soap film without obstructions produces an approximately uniform film thickness throughout the test section. \cite{Eshraghi2021} showed that thickness variations in the wake of a fixed circular cylinder are less than approximately 25\% of the average film thickness for $200\lesssim\text{Re}\lesssim300$. }
These thickness variations \added[id=R2]{in the wake} introduce a Boussinesq-type behavior, but the flow time scale \replaced[id=A]{is}{ was} fast enough that buoyancy effects  \replaced[id=A]{can be considered}{ were} negligible \citep{Auliel2015}. 
Thus these thickness variations, while crucial for the flow visualization, \replaced[id=A]{do}{ did} not play a significant role in the physics of the flow, which 
\replaced[id=A]{can be}{ was} assumed to 
approximate that of an incompressible fluid (see Section \ref{subsec:soapfilm}).  
\deleted[id=R2]{Figure \ref{fig:wakeFFT}(a) shows a sample of the flow visualizations obtained in our experiments.} 

\added[id=R1]{The oscillation of the cylinder caused some time-dependent, spatially-varying, out-of-plane displacement of the film. For sufficiently high amplitudes of oscillation frequency and amplitude, this deformation of the film prevented reflected light from reaching the camera, leaving dark areas in the interferometry images (see \cite{Masroor2021}).  Using an end-on view of the film, we determined  the maximum peak-to-peak out-of-plane displacement to be $\delta \approx 4$\,mm in the neighborhood of the cylinder for $A^*=0.94$ and $ 0.2 < f^* < 1.1$.  From these observations we estimated that   the out-of-plane displacement in this  system had a negligible effect on the wake dynamics in the regions of interest.  
}

Quantification of the wake frequencies was accomplished by examining the spatially-averaged, time-dependent pixel intensity (with value $0 \le \bar{I}\le  255$) from a limited portion of the imaged flow.  The interrogation window used for quantifying the fixed cylinder wakes, for example, was the rectangular domain measuring 1 pixel wide (in the streamwise direction) by 49 pixels high (in the transverse direction) shown as a white rectangle in figure~\ref{fig:wakeFFT}(a).  An example time trace of $\bar{I}(t)$ is shown in figure~\ref{fig:wakeFFT}(b). 
Variations in the wake that occurred on a time scale $T>1$\,s corresponded to flow features with a length scale of $\ell = U T \gtrsim 0.68$\,m, which was approximately the distance from the cylinder to the exit of the rectangular test section.  We thus applied a low-pass filter to the $\bar{I}(t)$ data in every case to exclude frequencies less than 1\,Hz.  
A Fast Fourier Transform (FFT) of the filtered $\bar{I}(t)$ data was then performed using MATLAB (MathWorks) to determine the corresponding power spectrum, such as shown in figure~\ref{fig:wakeFFT}(c). 
Every FFT in this manuscript was applied to the low-pass-filtered $\bar{I}(t)$ data taken from a specific interrogation window using the full experimental run, $0\le t \le 8.2$\,s, for that case.  

\subsection{Determining background flow speed}
\label{subsec:flowspeed}

The undisturbed background flow speed of the soap film was measured directly using Particle Tracking Velocimetry (PTV). For these measurements, the soap film solution was seeded during preparation with 10\,$\mu$m-diameter hollow glass spheres, and these spheres were illuminated in the film with an Elinchrom modeling lamp (EL23006).  Representative particles were identified in successive video frames using the Image Processing Toolbox in MATLAB (Mathworks).  Flow speed was calculated as the average distance traveled by these particles per unit time.  

\begin{figure}
	\centering
  \includegraphics[width=85mm]{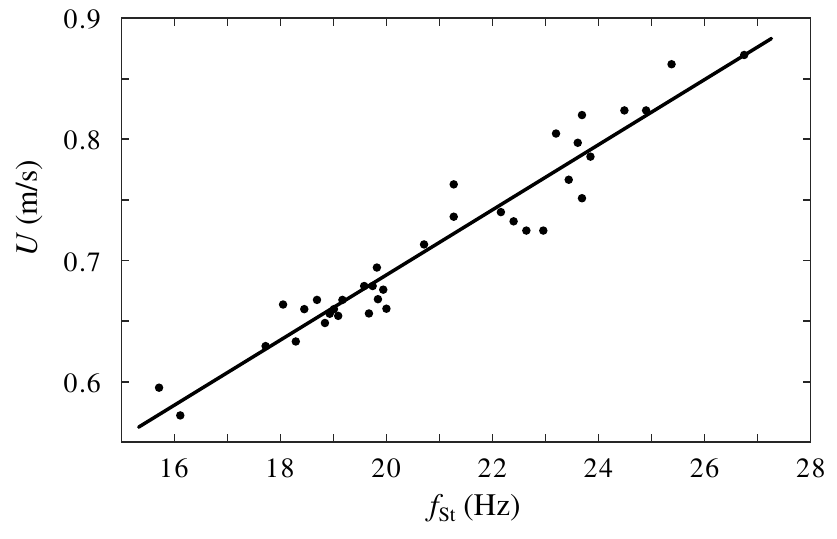}
  \caption{Variation of vortex shedding frequency, $\fst$, with background flow speed, $U$, in the soap film system for a fixed cylinder with diameter $D=6.35$\,mm.  Data is presented with $U$ as the dependent variable for analysis purposes.  Solid line is the linear data fit given in \eqref{eq:fvsU}. }
\label{fig:fvsU}
\end{figure}

We found that the seeding particles interfered with the interferometry flow visualization, making it impractical to directly measure the flow speed when conducting the oscillation experiments.  We thus correlated  flow speed with the vortex shedding frequency for a stationary cylinder with diameter $D=6.35$\,mm in our specific study. Multiple trials were conducted at various flow speeds in the range $0.55\,\text{m/s} < U < 0.90\,\text{m/s}$, with videos of the wake patterns recorded as described in section~\ref{subsec:viz}. 
The vortex shedding frequency, $\fst$, was identified here as the primary  peak in the FFT data, as described in section~\ref{subsec:viz}.  This approach to determining $\fst$ was validated through direct, manual calculations of vortex shedding frequency in a few select cases. 
The resulting relationship between $U$ and $\fst$ observed for this system is shown in figure~\ref{fig:fvsU} with $U$ taken as the dependent variable, which can be represented by the linear fit
\begin{equation}\label{eq:fvsU}
    U = (0.02686\,\text{m}) f_{\text{St}} + (0.1508\,\text{m/s})
\end{equation}
with correlation coefficient $R^2= 0.9281$.
Although this relationship is expected to be a function of channel inclination angle, channel aspect ratio, and soap solution composition, these parameters were held fixed for all of the experiments used in this study, and thus this relationship was assumed to hold for all cases presented here.

For the oscillation experiments, the background flow speed was estimated 
by first determining the shedding frequency with the cylinder held stationary and then using the relationship in \eqref{eq:fvsU}.  Flow speed was adjusted via the flow control valve to obtain a (fixed cylinder) shedding frequency for each consecutive set of experimental runs in the range $18.8 < \fst < 21.0$. Several successive video recordings were analyzed to determine a precise value of $\fst$ for each set of runs. The average shedding frequency for all experimental runs was $\fst = 19.58$\,Hz, which by equation~\eqref{eq:fvsU} corresponds to an average flow speed of  $U\approx 0.68$\,m/s.
\subsection{Determining the Reynolds number}
\label{subsec:determiningRe}
The standard approach to calculating Reynolds number requires an accurate value for the fluid viscosity.  In a flowing soap film, the viscosity value depends strongly on the thickness of the film;
\citet{Trapeznikov1957} has shown that the viscosity of a soap film can be represented as $\nu = \nu_w + 2 \nu_s/h$, where $\nu_w$ is the bulk viscosity of water, $\nu_s$ is the viscosity of the surfactant-dominated free-surface layers, and $h$ is the thickness of the film \citep[see also][]{Vorobieff1999}. For the very thin films generated in soap film experiments, the thickness is difficult to measure precisely but plays an important role in determining viscosity.  

\cite{Gharib1989} were the first to suggest that the Reynolds number for the flow past a cylinder in a flowing soap film could be estimated by observing the Strouhal number and then estimating the Reynolds number using the empirically-known relationship between Re and St.  
This indirect approach to determining the Reynolds number for flow past a circular cylinder in a flowing soap film has since been used by several investigators; see, e.g., \cite{Vorobieff1999}, \cite{Wen2001}, \cite{Jia2007}, and \cite{Wang2010}.
We followed this approach for estimating the Reynolds number in our system. 
We first used the observed vortex-shedding frequency, $\fst$, the corresponding flow speed of the soap film $U$ from \eqref{eq:fvsU}, and the cylinder's diameter $D$ to calculate the Strouhal number, $\text{St} = \fst D/U$. 
We then used the St-Re relationship from \cite{Williamson1998},
\begin{equation}
    \text{St} = 0.285 - \frac{1.390}{\sqrt{\text{Re}}} + \frac{1.806}{\text{Re}},
\label{St-Re}
\end{equation}
to determine the Reynolds number.  
The range of vortex-shedding frequencies observed in our experiments corresponds to an average Reynolds number of $\text{Re} = 235 \pm 14$.

\subsection{Cylinder oscillation}
\label{subsec:oscillation}

As noted above, an acrylic rod with diameter $D=6.35$\,mm was mounted on a plate oriented parallel to the test section of the soap film channel, so that the cylinder penetrated the soap film at a 90$^{\circ}$ angle.  For the oscillation experiments, the plate was attached to the shaft of a motor via a Scotch-yoke mechanism that converted the circular motion of the shaft to sinusoidal translation of the cylinder. The plate and motor were mounted to a vibration-isolation table, and care was taken to ensure that the frame supporting the soap film did not come into contact with the motor mount or the vibration-isolation table.

\begin{figure}
	\centering
	\includegraphics[scale=0.9]{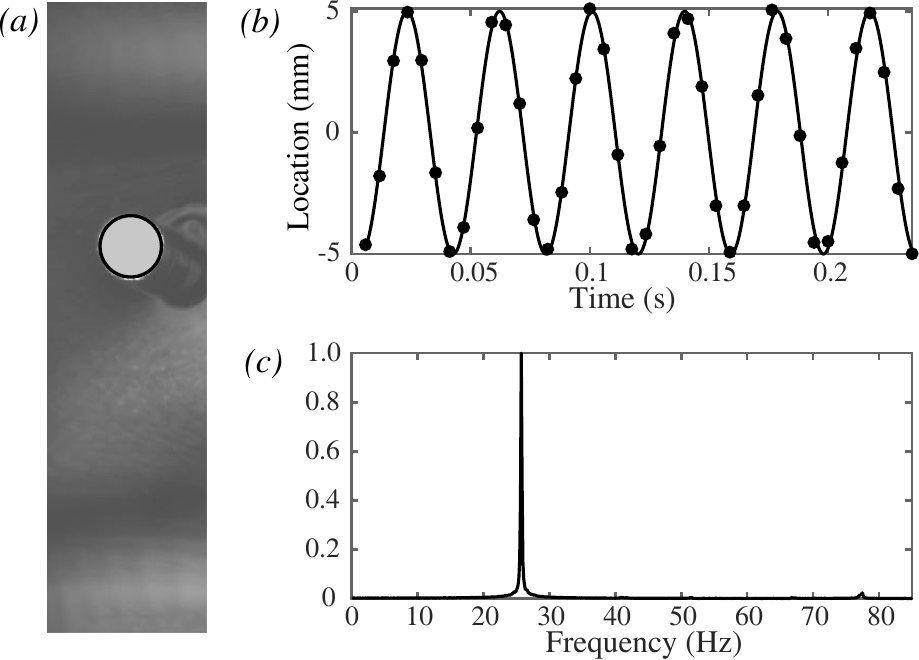}
	\caption{(a) Representative location of the circle identified with the Hough Transform superimposed on a cropped frame from the video; fluid flow is from left to right. (b) Example time-trace of the vertical position of the cylinder, superimposed on a sine curve with frequency 25.74 Hz. (c) Power spectrum of the example time-series data using a Fast Fourier Transform, showing a clear peak at 25.74 Hz.}
	\label{fig:CylinderFFT}
\end{figure}

The cylinder's oscillation frequency, $f$, was verified using image analysis of the recorded videos as described in section~\ref{subsec:viz}.  
We extracted from each video frame a small spatial region (150 \by 600 pixels) within which the motion of the cylinder was contained, as shown in figure~\ref{fig:CylinderFFT}(a). A Hough transform was performed on this frame region to track the position of the cylinder, giving a time trace of the cylinder such as illustrated by the solid circles in figure~\ref{fig:CylinderFFT}(b). We then used a Fast Fourier Transform (FFT) on this time series to extract a power spectrum, which in every case had a single well-defined peak at the oscillation frequency like that shown in figure~\ref{fig:CylinderFFT}(c). 

\begin{figure}
	\centering
		\includegraphics[scale=1]{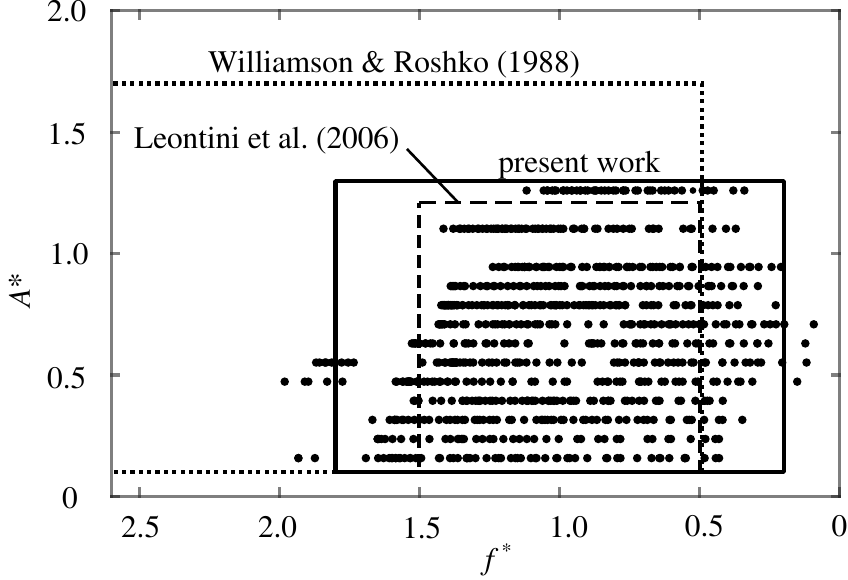}
	\caption{Comparison of the range of $\faspace$ parameter space explored in this work and in previous studies.
	A total of 840 experiments were conducted; each dot represents a single experiment at a particular (non-dimensional) amplitude, $A^*$, and oscillation frequency, $f^*$. 
	Note that $f^*$ increases to the left, a convention used to facilitate comparison with previous work.	The range reproduced in figure~\ref{fig:map} is indicated by a solid rectangle, the dashed rectangle shows the region of parameter space explored by \cite{Leontini2006}, and the dotted rectangle shows the (approximate) region of parameter space covered by \cite{Williamson1988}.}
	\label{fig:unlabeledmap}
\end{figure}

The frequency-amplitude values used for the experiments reported here are shown in figure~\ref{fig:unlabeledmap}.
The amplitude of the cylinder's oscillation was determined by the offset of the cylinder's axis from the axis of the motor's rotation, which for a given Scotch yoke mechanism is fixed. Therefore, we varied the amplitude of oscillation by fabricating yokes with different eccentricities, for a total of thirteen different amplitudes $A\in\{$1.0\,mm, 1.5\,mm, 2.0\,mm, 2.5\,mm, 3.0\,mm, 3.5\,mm, 4.0\,mm, 4.5\,mm, 5.0\,mm, 5.5\,mm, 6.0\,mm, 7.0\,mm, 8.0\,mm$\}$.
The frequency of oscillation was varied along a continuum for each amplitude by varying the amount of power provided to the motor from case to case. In these experiments, the frequency of oscillation was kept in the range $f < 40\,\text{Hz}$ to \replaced[id=R1]{avoid the}{ minimize} out-of-plane vibrations of the soap film that \replaced[id=R1]{interfered with the interferometry}{ were produced} when exciting at high frequencies, especially at large amplitudes. 
The relationship between our experiments and the region of $\faspace$ parameter space explored by previous studies is also shown in figure~\ref{fig:unlabeledmap}. 
Note that the comparison with the parameter space studied by \cite{Williamson1988} is only approximate, since they report results in terms of the reduced velocity $U^* = \text{St}/f^*$ and the Strouhal number varies with Reynolds number for this data.

The position of the cylinder in its motion can be characterized by a dimensionless time, $t^* = t/\tau = t f$.  We chose to define $t^*=0$ when the cylinder was located at the centerline of its motion and moving in the downstroke.  Throughout this manuscript we restrict $-0.5< t^* < 2.0$ and use this variable to represent the relative position of the cylinder in its cycle. 

\subsection{A note on cylinder diameter}
\label{subsec:diameter}

The cylinder was surrounded by a meniscus at the cross-section where it intersected with the soap film, and this meniscus moved with the cylinder as it oscillated.  \cite{Couder1986} proposed calculating an effective diameter of the cylinder in such a case by comparing the vortex shedding of the fixed cylinder with that of a thin fixed disk aligned with the plane of the film.  The resulting (larger) effective diameter is then used in place of the physical diameter in all calculations.  

We estimated an effective diameter for our system using the following procedure. We conducted fixed cylinder wake experiments using a very thin circular disk made from clear polyester film (McMaster--Carr, USA) with diameter $D=6.35$\,mm and thickness $\diskthickness = 25.4\,\mu$m.  The disk was held in the plane of the film by surface tension. Since the disk thickness was close to the film thickness, the meniscus effect was assumed to be negligible. Using the St–Re relationship from \eqref{St-Re}, the effective viscosity was estimated to be $\nu_{\text{eff}} = 20.1 \pm 0.3$\,mm$^2$/s. The relationships between $U$, $\fst$, St and Re then give a non-linear system of equations for the unknown effective diameter, $D_{\text{eff}}$, which we solved using an iterative process to find $D_{\text{eff}}\approx7$\,mm.  

Replacing the cylinder diameter, $D=6.35$\,mm, in our calculations with the effective diameter, $D_{\text{eff}}=7$\,mm, increases the estimated Reynolds number and decreases the dimensionless oscillation amplitude, $A^*$.  This change does not \replaced[id=R3]{affect}{ effect} the interpretation of our results, except as it impacts direct comparisons with the results of others.  Since our experimental parameters are not an exact match with any other work, and since there is some subjectivity in the interpretation of the effective diameter, particularly for a moving cylinder, we have chosen to report all results in terms of the physical cylinder diameter, $D$.

\section{Results}
\label{sec:results}

\begin{figure}
	\centering
		\includegraphics[scale=1=]{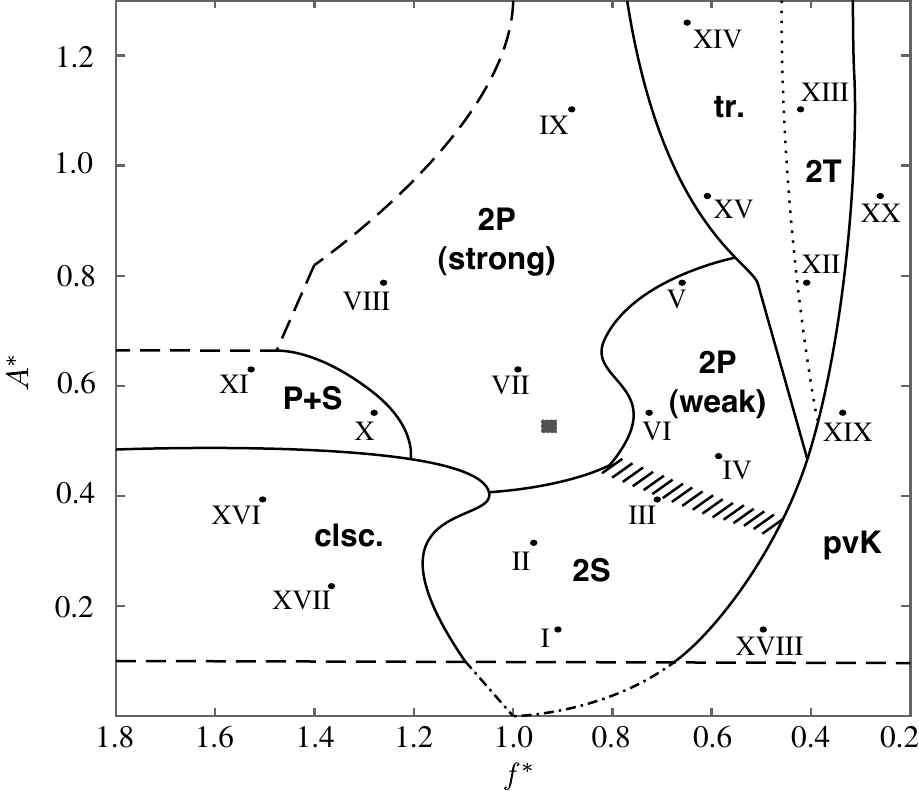}
	\caption{\replaced[id=A]{Wake-mode map delineating regimes }{Bifurcation diagram} of flow structure in the wake of a cylinder oscillating in a flowing soap film at $\text{Re}\approx 235$. Solid lines indicate \deleted[id=A]{clearly-discernible} boundaries between distinct regions of synchronization; \deleted[id=R2]{dashed and dotted lines indicate boundaries that were determined with a lower level of precision, as described in the text.} \added[id=R2]{dashed lines mark limits on data collection; boundaries marked by hatched, dotted, and dashed-dotted lines are described in the text.}
	Points labeled with \deleted[id=A]{uppercase} Roman numerals represent specific experiments referenced in the text.
	The filled square marks the location in parameter space for the example considered in figure \ref{fig:2P_detail}.
	Regions are labeled according to the dominant wake pattern as coalescing (clsc), 2S, perturbed von \karman\ (pvK), P+S, 2P (weak and strong), transitional (tr), and 2T.  The inverted axis for $f^*$ is used to match convention, as decreasing $f^*$ corresponds to increasing reduced velocity, $U^*$.}
	\label{fig:map}
\end{figure}

Over a large portion of the parameter space explored in this study, coherent vortices formed in the wake of an oscillating cylinder in patterns that were observed to repeat periodically and synchronously with the cylinder motion.  
In figure \ref{fig:map} we show the regions of parameter space where we observed the same vortex shedding patterns to occur consistently.  
We follow the terminology of \cite{Williamson1988} and \cite{Williamson2004a} by denoting a single vortex with `S', a pair of vortices with `P', and a triplet of vortices with `T'. The frequency axis decreases from left to right in accordance with the convention used in the literature for controlled-oscillation experiments and simulations.

The boundaries shown in figure~\ref{fig:map} do not mark abrupt changes in the wake structure.  Rather, they denote the approximate locations of ``transition zones'' over which the wake structure transitions from one type to another.  We speculate that this smooth shift in wake type is related to the topological bifurcation that occurs as the wake of a stationary cylinder transitions from a time-periodic wake to the von~\karman\ vortex street \citep{heil2017jfm}.

The \textit{primary synchronization} region --- in the neighborhood of $f^* \approx 1$ --- was found to be dominated by 2S wakes at low amplitudes and 2P wakes at higher amplitudes. For the mid-range amplitudes in the regime identified as `(weak) 2P', one vortex in each pair is considerably weaker than its partner.
As described in section~\ref{section:2P}, the weak 2P mode is not clearly distinct from the 2S mode, and the approximate boundary between these two regimes is shown as a \replaced[id=R2]{hatched region}{ dotted line} in figure~\ref{fig:map}.
At the high frequencies and small amplitudes in the coalescing (clsc) region, multiple vortices are shed from the cylinder per oscillation, and these vortices quickly merge to form a secondary street.  At high frequencies and large amplitudes (i.e., the empty upper-left portion of figure~\ref{fig:map}), \deleted[id=R1]{significant} out-of-plane vibrations in the soap film \replaced[id=R1]{interfered with visualization}{ prevented exploration} of the wake structure\added[id=R1]{, so this region was excluded from our analysis}. 
The edges of this region are dictated by limitations in our experimental capabilities, not a known change in wake structure, so this region is bounded by dashed lines.  
At low frequencies and small amplitudes, the 2S and (weak) 2P structures transition to what we refer to as a `perturbed von \karman' (pvK) wake, in which the vortex formation is not locked on to the cylinder's motion; the pvK pattern does not appear to be discussed in the existing literature. Finally, for large amplitudes and decreasing frequencies, the wake patterns consisted progressively of a transitional mode in which \replaced[id=A]{complex or unclear}{ or no clear periodic} wake structure\replaced[id=A]{s  were}{ was} observed, a `2T' mode with two triplets of vortices shed during each cycle of the cylinder's motion, and the pvK mode.  More highly resolved data would be needed to resolve the tr-2T boundary more accurately, and thus we show this approximate boundary as a \replaced[id=A]{dotted}{dashed} line. \added[id=A]{No data were collected below the dashed line at $A^* = 0.1$. The extension of the 2S region's boundaries to $(f^*, A^*) = (1,0)$ is based on the fact that a stationary cylinder will shed vortices only at $f=\fst$ and the assumption that in our system a small amplitude oscillation $A^*=O(\epsilon)$ will produce vortex shedding with $f^*\approx 1$.  
This extension is similar to that of \cite{Leontini2006}; see  figure~\ref{fig:WR-L}(b).  
However, this extension is purely speculative, and in a three-dimensional system a critical oscillation amplitude of $A^*\gtrsim0.1$ is needed to achieve `lock on' \citep{Koopmann1967}.
}
In the following sections, representative examples are discussed from each region of $\faspace$ space identified in figure~\ref{fig:map}.
\added[id=A]{The experimental videos for each of the twenty examples enumerated in figure~\ref{fig:map} and discussed below are documented in \cite{Masroor2021}.}

\begin{figure}
	\centering
		\includegraphics[scale=1]{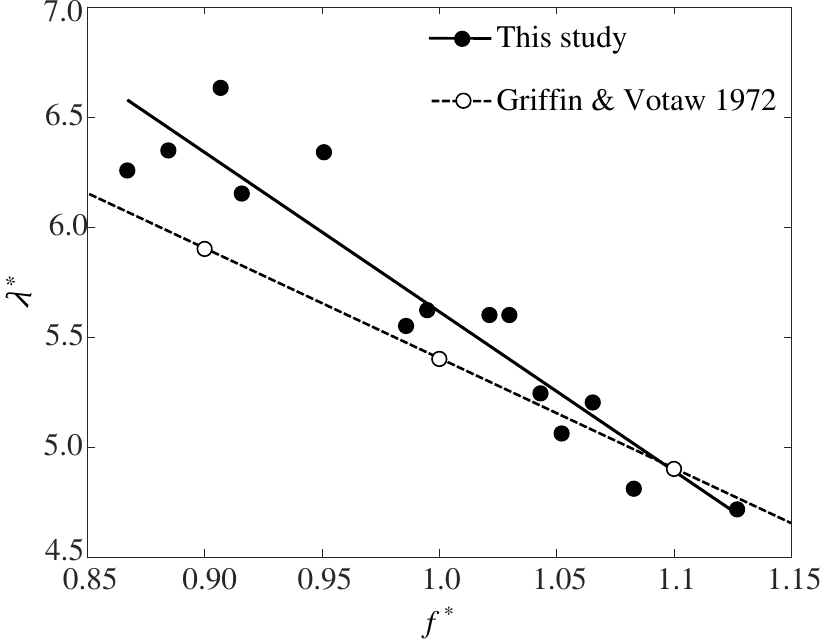}
	\caption{Variation of normalized wake wavelength, $\lambda^*$, 
	with increasing normalized cylinder oscillation frequency, $f^*$, for a fixed value of oscillation amplitude, $A^*$, and Reynolds number. \replaced[id=A]{Open}{ Closed} circles are from \cite{Griffin1972} at $A^* = 0.30$ for $\text{Re}=144$; \replaced[id=A]{dashed}{ solid} line is linear fit to this data. \replaced[id=A]{Closed}{ Open} circles are from the present study at $A^* = 0.315$ and $\text{Re}=235$, corresponding to a horizontal slice through the 2S region in  figure~\ref{fig:map}; \replaced[id=A]{solid}{ dashed} line shows a linear fit to this data.}
	\label{fig:wavelength}
\end{figure}

At any given amplitude of oscillation, the wake wavelength was found to decrease with increasing oscillation frequency. This dependence of wake wavelength on cylinder oscillation frequency is broadly consistent with the observations of \cite{Griffin1972}, and a comparison for $A^* \approx 0.3$ is shown in figure~\ref{fig:wavelength}.  The good agreement that is shown despite the differences in Reynolds number and experimental method supports the validity of using a flowing soap film system for quantitative descriptions of two-dimensional wakes.


\subsection{2S mode}\label{section:2S}

\begin{figure}
\centering
  \includegraphics[scale=0.9]{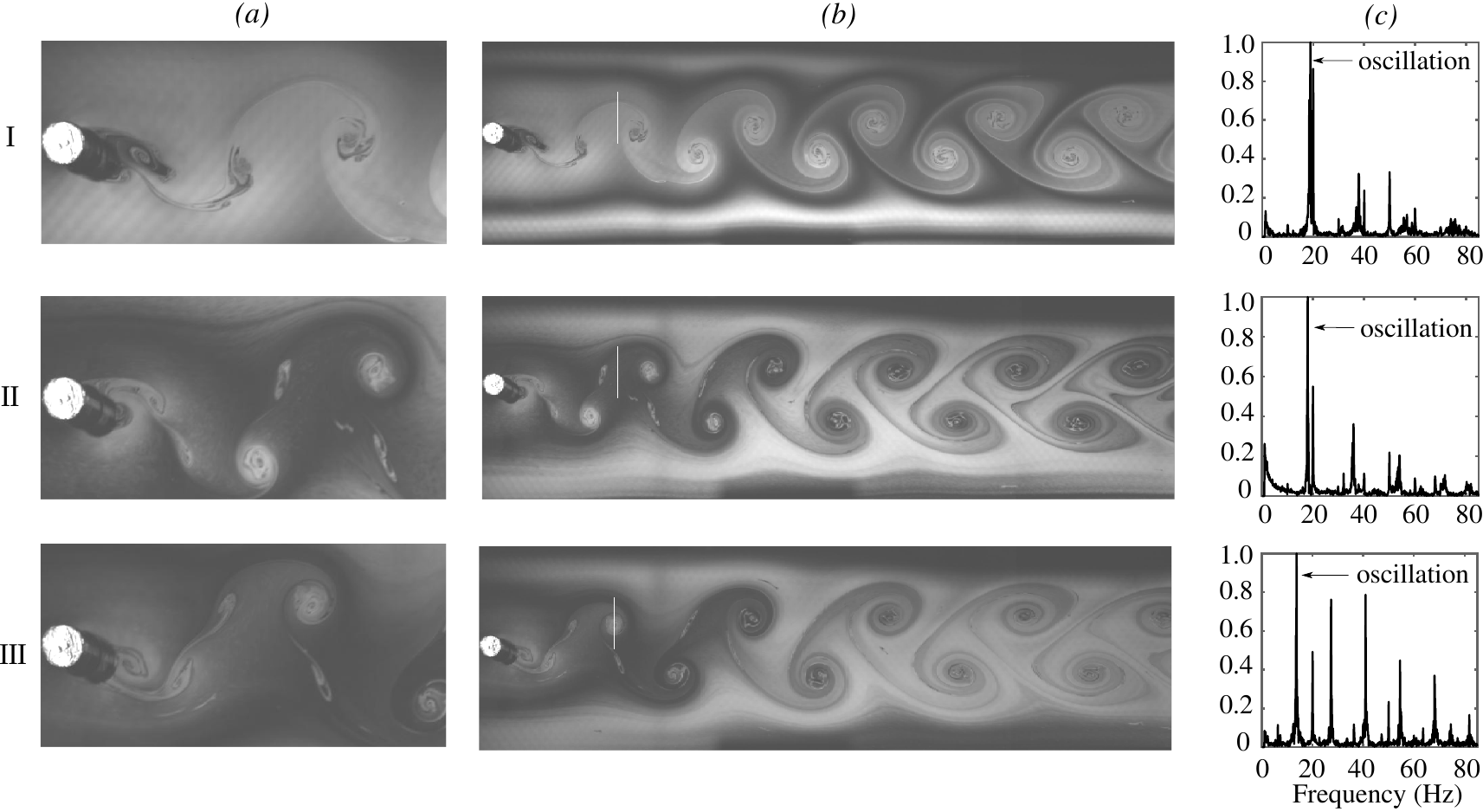}
  \caption{Representative cases of 2S vortex wakes.  Rows are labeled according to case with parameters I: $(f^*, A^*) \approx (0.90, 1.16)$ at $t^* = 0.578$, II: $(f^*, A^*) \approx (0.95, 0.32)$ at $t^* = 0.728$, and III: $(f^*, A^*) \approx (0.70, 0.39)$ at $t^* = 0.366$; cases are also identified with the same labels in figure~\ref{fig:map}.  Columns show (a)~the near- and mid-wake regions (image size: 600 \by 300 pixels $\approx 10D \by 5D$), (b)~the extended wake region (image size: 2053 \by 600 pixels $\approx 34.2D \by 10D$), and (c) the normalized power spectrum for each case, with the frequency corresponding to cylinder oscillation identified.  White rectangles in column~(b) show interrogation window (1 \by 150 pixels) used to calculate power spectrum. Contrast has been enhanced in wake images to improve clarity.}
\label{fig:2S}
\end{figure}

The low-amplitude region of parameter space at oscillation frequencies close to the natural frequency of vortex shedding was found to be populated by synchronized 2S wakes, in which two vortices of equal and oppositely-signed strengths are formed from the vorticity shed during each half-cycle of cylinder motion. The clockwise vortex is shed during the cylinder's upstroke (for left-to-right flow), and the counter-clockwise vortex is formed from vorticity shed during the downstroke. Three example cases, for the parameter values labeled I--III in figure~\ref{fig:map}, are shown in figure~\ref{fig:2S}.
Synchronized 2S wakes such as these have been widely reported to be observed in the vicinity of the low-amplitude primary resonance in studies of both controlled \citep{Williamson1988,Govardhan2000,Leontini2006} and free oscillations \citep{Khalak1999}. Typically, the power spectra from these wakes are strongest at the oscillation frequency 
despite the low amplitude of oscillation, which suggests that these wakes are strongly locked-on to the cylinder oscillation. These power spectra, such as shown in figure~\ref{fig:2S}(c), were determined for our examples by taking an FFT of the average image intensity in a 1 \by 150 pixel rectangle located a distance $7.2D$ downstream of the cylinder (see figure~\ref{fig:2S}(b)); thus, the cylinder oscillation frequency can appear in the power spectra only through its influence on the cylinder wake.  

The synchronized 2S mode of vortex shedding in the wake of an oscillating cylinder is physically quite distinct from what is known as the von \karman\ mode of vortex shedding in the wake of a \emph{fixed} cylinder. Although the resulting flow looks quite similar (compare figure \ref{fig:2S}(b) with figure \ref{fig:wakeFFT}), the 2S mode occurs due to \replaced[id=A]{an}{ a fortuitous} interplay between the motion of the cylinder and the motion of the fluid, with the cylinder's motion causing the shear layers from each side of the cylinder to roll up into exactly two counter-rotating eddies during each period of the cylinder's motion. In the von \karman\ wake, there is no motion of the cylinder, and the shear layers independently arrange themselves into a series of counter-rotating eddies at the Strouhal frequency. 
In other words, the fixed-cylinder vortex wake formation is a `shear layer instability' phenomenon \citep{heil2017jfm}, while the 2S wake of an oscillating cylinder \replaced[id=R3]{relies also on a}{ is the result of a more involved} `resonance' phenomenon.

Some variation in the detailed structure of the near wake was observed within the region labeled 2S. When $f^* \approx 1$ and $A^*$ was low (see, for example, case~I in figure~\ref{fig:2S}), the clockwise and counterclockwise shear layers from the top and bottom of the cylinder rolled up discretely into single vortices of opposite sense, resulting in a `clean' structure in the far wake that strongly resembles the von \karman\ street.  Increasing the amplitude (e.g., case~II) and also reducing the oscillation frequency (e.g., case~III) resulted in each shear layer splitting into a weak and a strong vortex of opposite sense, giving two oppositely-signed pairs of vortices in the near wake for each period of oscillation. However, in the 2S region of figure~\ref{fig:map}, this appearance of a weak 2P structure immediately after the cylinder was short-lived, and within two periods of motion 
the wake assumed a clear 2S structure. This near-wake behavior appears to be consistent with that shown in corresponding 2D computational results \citep[figure 7]{Leontini2006}, in which each newly-shed vortex had a  weak `tail' that was not counted when classifying the wake. In section~\ref{section:2P} we show  that a modified form of this same process was responsible for the formation of 2P wakes.

\subsection{2P mode}\label{section:2P}

The largest portion of the primary synchronization region was found to exhibit the 2P mode of vortex shedding, in which two pairs of oppositely-signed vortices were shed per oscillation cycle. This wake mode was  observed primarily in the high-amplitude range ($0.4 < A^* <1.3$) of the primary synchronization region ($f^*\approx 1$), which is similar to the results reported by  \cite{Williamson1988} for $300 < Re < 1000$.
In contrast, \cite{Leontini2006} did not find any 2P wakes in their two-dimensional computations.  
Two main types of 2P wake were identified in our experiments: a `weak 2P' mode and a `strong 2P' mode that are distinguished primarily by the structure and relative strength of the pairs in the mid- and far- wake. A similar weak 2P mode has been identified in other studies \cite[see, e.g., the $2\text{P}_o$ mode in figure 4 of][]{Morse2009} when spatially-resolved data in the near wake have been available.

\subsubsection{Weak 2P mode}

\begin{figure}
\centering
  \includegraphics[scale=0.9]{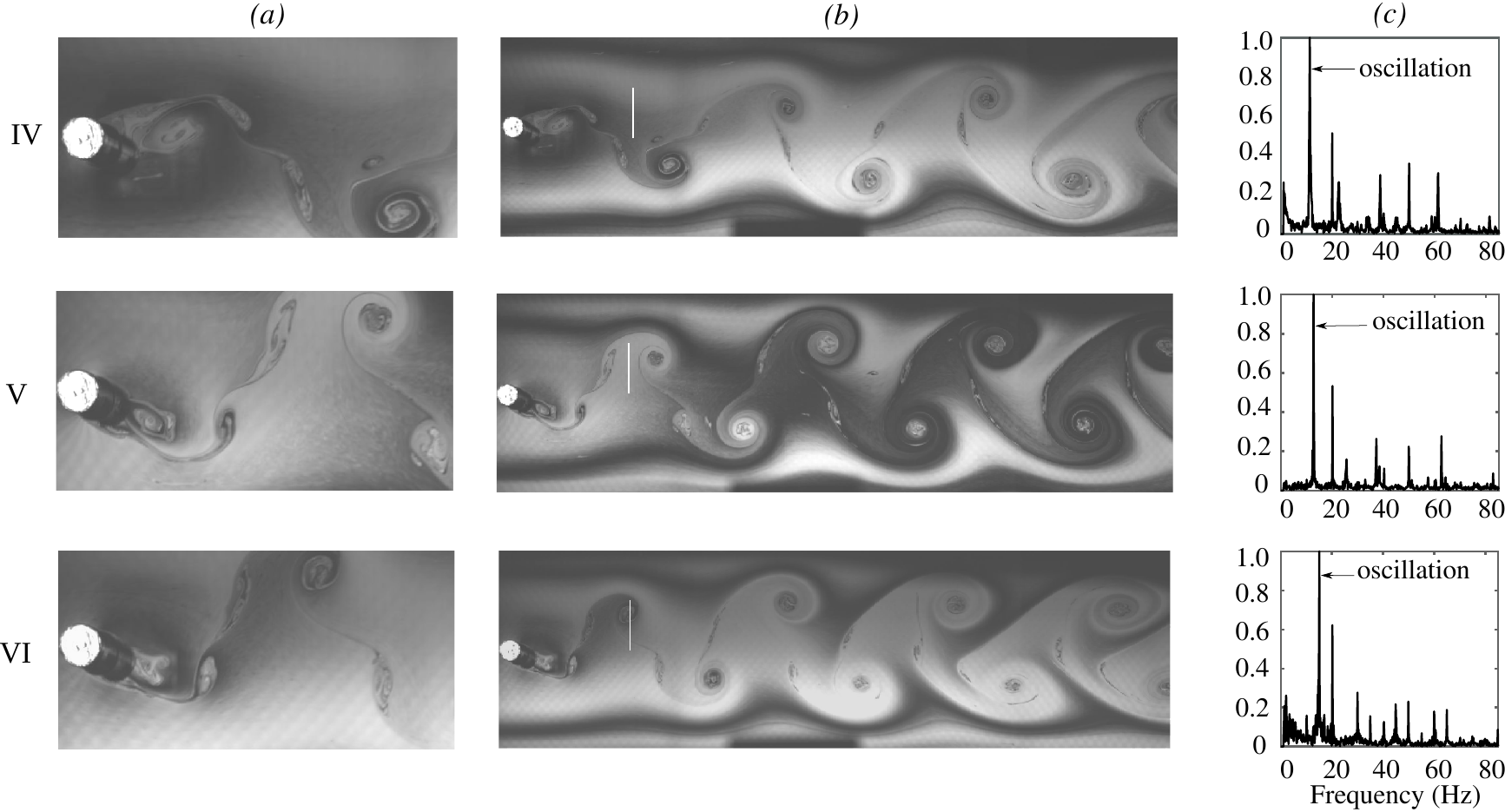}
  \caption{Representative cases of weak 2P vortex wakes.  Rows are labeled according to case with parameters IV: $(f^*, A^*) \approx (0.57, 0.47)$ at $t^* = 0.852$, V: $(f^*, A^*) \approx (0.65, 0.79)$ at $t^* = 0.366$, and VI: $(f^*, A^*) \approx (0.72, 0.55)$ at $t^* = 0.339$; cases are identified with the same labels in figure~\ref{fig:map}.  Columns show (a)~the near- and mid-wake regions (image size: 600 \by 300 pixels $\approx 10D \by 5D$), (b)~the extended wake region (image size: 2053 \by 600 pixels $\approx 34.2D \by 10D$) with interrogation window shown in white, and (c) the normalized power spectrum for each case.  Contrast has been enhanced in the wake images to improve clarity. }
\label{fig:weak2P}
\end{figure}

A `weak 2P' mode was observed in the low-frequency, mid-amplitude region of the parameter space, as shown in figure~\ref{fig:map}. Three examples of these weak 2P wakes are shown in figure \ref{fig:weak2P}. 
Four coherent vortex structures are shed during each cylinder oscillation cycle.  The weaker, secondary vortices survive the roll-up process to a greater extent than in the 2S region, leading to a mid- and far-wake structure that differs from the von~\karman\ street. 
Each secondary vortex, although shed as its own coherent structure, is significantly strained by the stronger neighboring vortices  as it moves to a location mid-way between these successive strong vortices.

As described at the beginning of section~\ref{sec:results}, changes between regimes occur continuously over a (narrow) range of parameter space rather than abruptly at a distinct point. For the transition between the 2S and weak 2P modes, this change occurs over a broader range of parameter space than for the other transitions.   
A certain subjectivity in the distinction between the 2S mode and the low-amplitude weak 2P mode, therefore, is unavoidable, and we have denoted our estimate of the boundary between these two regions by a \replaced[id=R2]{hatched region}{ dotted line} in figure~\ref{fig:map}.

The weak 2P region shows a significant degree of variation in the wake details across the corresponding region of parameter space. 
The far-wake structure for case~IV --- which is close to the boundary with the 2S region --- is not very different from that of case~III (see figure~\ref{fig:2S}).
At higher amplitudes and lower frequencies, placing the system farther away from the 2S region of parameter space, a more distinct `weak 2P' structure emerges in the far wake, with the weaker member of the pair persisting for longer convective times (compare cases~V and VI with case~IV). However, in all parts of the weak 2P region the weaker vortex does eventually distort into a braid shear layer between successive strong vortices, with the process happening further downstream with greater distance (in parameter space) from the 2S region. 
The fact that this persistence of the weak vortices increases with decreasing frequency suggests that the transition from 2S to weak 2P wakes is a result of the slower transverse motion of the cylinder, which produces a greater distance between successive strong vortices and reduces their inductive effect on the weak vortices.

\subsubsection{Strong 2P mode}

\begin{figure}
\centering
  \includegraphics[scale=0.9]{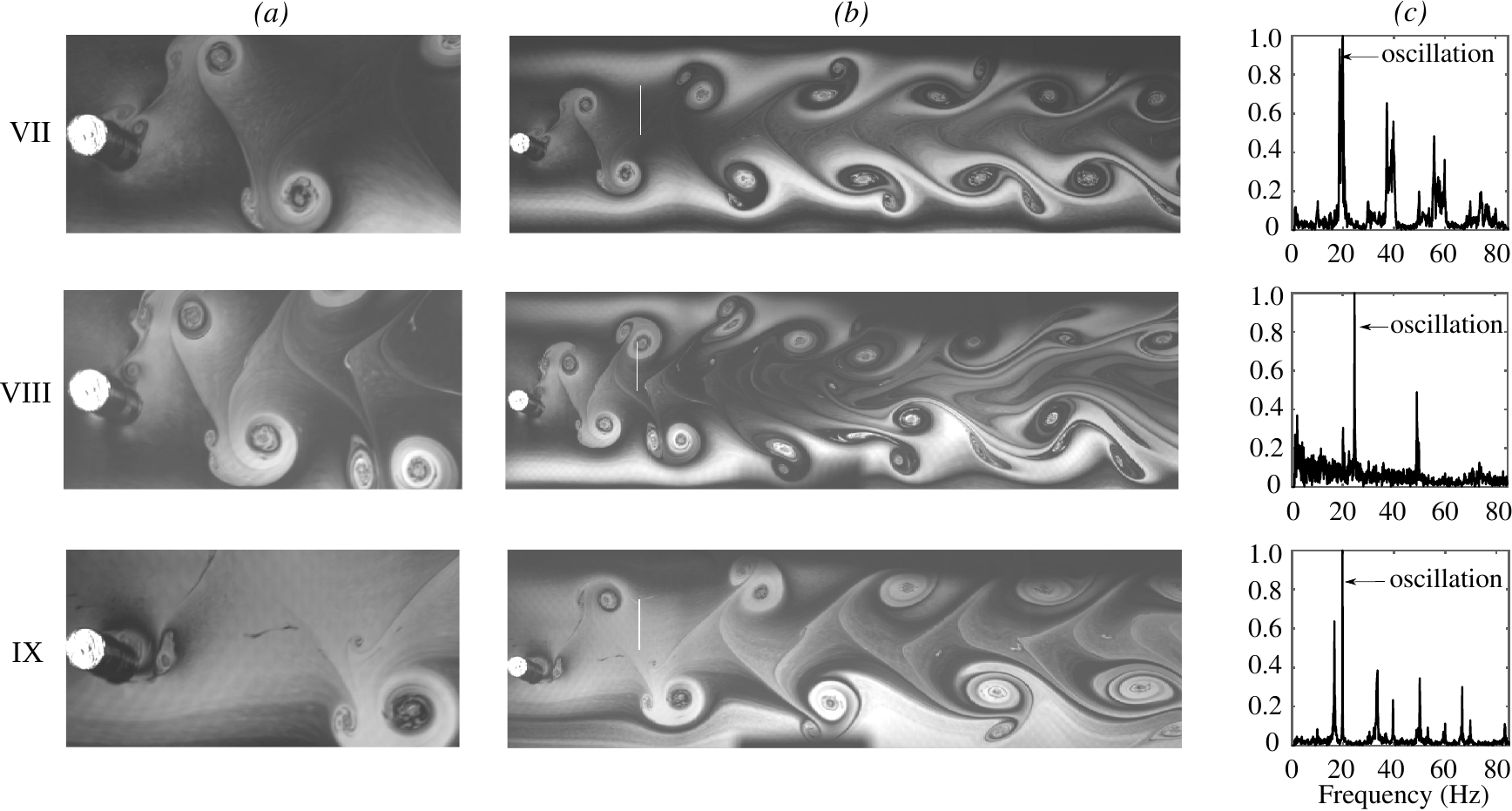}
  \caption{Representative cases of strong 2P vortex wakes.  Rows are labeled according to case with parameters VII: $(f^*, A^*) \approx (0.98, 0.63)$ at $t^* = 0.048$, VIII: $(f^*, A^*) \approx (1.26, 0.79)$ at $t^* = 0.967$, and IX: $(f^*, A^*) \approx (0.87, 1.10)$ at $t^* = 0.179$; cases are also identified with the same labels in figure~\ref{fig:map}.  Columns show (a)~the near- and mid-wake regions (image size: 600 \by 300 pixels $\approx 10D \by 5D$), (b)~the extended wake region (image size: 2053 \by 600 pixels $\approx 34.2D \by 10D$), and (c) the normalized power spectrum for each case.  Contrast has been enhanced in the wake images to improve clarity.}
\label{fig:strong2P}
\end{figure}

The second type of 2P wake --- which we refer to as the `strong 2P' mode --- occurs at higher amplitudes and oscillation frequencies relative to the weak 2P mode, as shown in figure~\ref{fig:map}. We identify this mode as distinct from the weak 2P mode because the wake structure in this strong mode consists clearly of vortex pairs well into the far wake region, as shown in figure~\ref{fig:strong2P} for three representative examples.  
The vortices in each pair were of noticeably unequal strength throughout the entire parameter space, even in this strong 2P region.  However, in the strong 2P mode the strength of the weaker vortex and its distance from the paired stronger vortex remain large enough to prevent the weaker vortex from being strained into a braid shear layer in the mid-wake region.  

The relative motion of vortices in each of cases~VII and IX is reminiscent of the `orbiting' and `exchanging' modes identified by \cite{Basu2015} in the motion of two pairs of point vortices with glide-reflective symmetry in a periodic strip. 
Due to the near-periodicity of the 2P wake (in most cases), the relative vortex positions in successive downstream pairs can be interpreted as representing the stroboscopic time evolution of a single pair.  
In case~VII, the weaker vortex of each pair orbits the stronger vortex for less than one full revolution when it appears to exchange partners with the strong vortex in the neighboring (downstream) pair just as it leaves the frame, giving rise to relative vortex motion in an `exchanging' mode.  In case~IX, the weaker vortex orbits the stronger vortex and remains in close proximity to its original partner throughout the visible wake, showing behavior in an `orbiting' mode.  This progression from an exchanging mode to an orbiting mode for increasing $A^*$ with $f^*\approx 1$ is consistent with the relative vortex strengths in each pair becoming more similar as the oscillation amplitude is increased.  


\subsubsection{2P wake formation}

\begin{figure}
	\centering
  \includegraphics[scale=0.9]{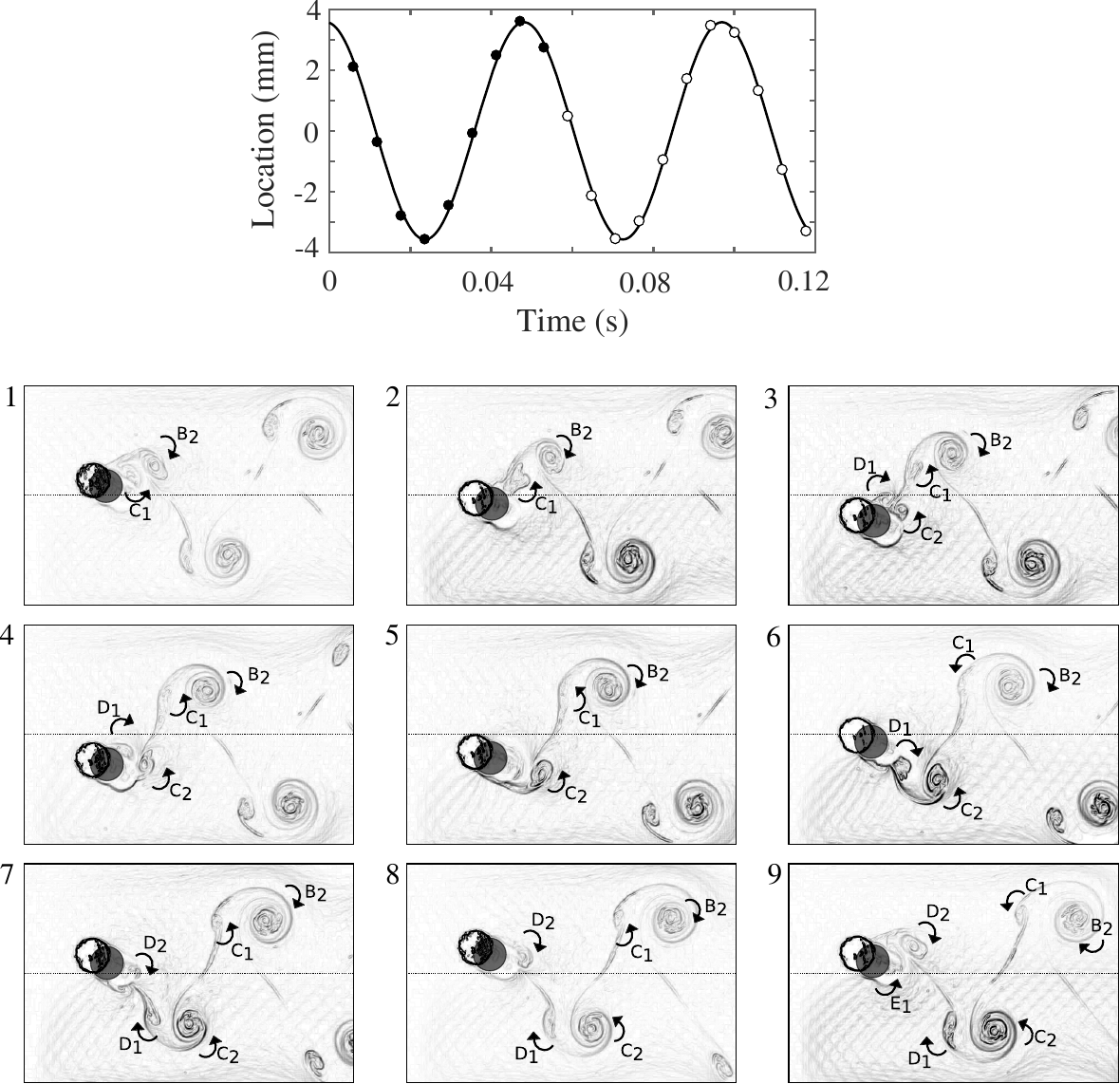}
  \caption{Vortex formation process for the strong 2P mode wake with $(f^*,A^*) \approx (0.92,0.55)$; corresponding point in parameter space is marked by a solid square in figure~\ref{fig:map}. Top panel: Cylinder positions in time (solid circles) corresponding to the panel images; solid line shows \added[id=A]{a} sine curve fit to \replaced[id=A]{all observed positions over the full 8.2-second run, not just the points shown here}{the observed trajectory}.  
  Numbered panels: flow structure for {$t^*_i = \{ -0.107, 0.007, 0.120, 0.234, 0.348, 0.462, 0.576, 0.689, 0.803, 0.917 \}$}, with $i$ given by the panel label. Each wake image measures 599 \by 398 pixels $\approx 10D \by 6.6D$). Cylinder cross-section in contact with the soap film has been marked by a filled circle. Images have been processed to highlight flow structure.  Vortex labels are discussed in the text. Vortex $D_1$ is obscured from view in the fifth frame due to the parallax effect and thus is not labeled there. \replaced[id=A]{Compare with}{ Cf.}~figure~11 in \citet{Williamson1988}.   }
\label{fig:2P_detail}
\end{figure}

In order to better understand the process of vortex shedding in the strong 2P mode, which dominated the primary synchronization region in our experiments, nine successive video frames of the cylinder and its near wake were analyzed for an oscillation  amplitude $A=3.5$\,mm ($A^*=0.55)$ and frequency $f=19.34$\,Hz.  For this case, $\fst=21.0$, giving $f^* = 0.92$.  The point $(f^*, A^*) = (0.92, 0.55)$ is marked by a filled square in figure~\ref{fig:map}. Additional processing to highlight the flow structure gives the images shown in figure~\ref{fig:2P_detail}, which the reader is encouraged to compare with figure 11 of \cite{Williamson1988}.

The nine frames capture approximately one full cycle of the cylinder's motion, comprising one period of the sine curve traced by the cylinder in a frame moving with the free stream, as illustrated in figure~\ref{fig:2P_detail} (top panel). Due to a parallax effect, the top of the cylinder, which appears clearly in the images, does not coincide with the location at which the circular cross-section intersects the soap film; a solid circle has been added in each frame to mark this location.  The periodic nature of vortex shedding can be verified by the similarity between frames~1 and 9, which are at approximately the same phase of the cyclic motion. Each vortex is identified by an alphanumeric label, with syntax chosen to mirror that in \cite{Williamson1988}. 

Figure \ref{fig:2P_detail} shows the cylinder starting near the beginning of its downstroke. Vortex $B_2$ has already been formed, while vortex $C_1$ is observed shedding and forming over the first two frames as the cylinder moves downward. Then, under the influence of its stronger neighboring vortex $B_2$, vortex $C_1$ is rapidly pulled away from the cylinder between the second and third frames. Thus paired together, these two vortices subsequently convect away from the cylinder without much relative motion.

The shedding and convecting away of the counterclockwise vortex $C_1$ relatively early in the cylinder downstroke allows enough time for another counterclockwise region of vorticity to develop on the cylinder's edge during the remainder of the downstroke; the beginnings of this development can be seen at time $t^*_3$. This new vortex, labeled $C_2$, is then shed from the cylinder not due to the induction of a neighboring vortex (as was $C_1$) but due to the `jerk' of the cylinder during the turning point in its sinusoidal motion, which occurs between the fourth and fifth frames. In later frames, it can be seen that $C_2$ is a much stronger counterclockwise vortex than $C_1$, and these two vortices remain connected to each other via a braid shear layer with like-signed vorticity owing to them being  
shed during the same downstroke.

During the cylinder's upstroke, clockwise vorticity is shed. The incipient clockwise vorticity (labeled $D_1$) can already be seen developing in the third frame as the cylinder's downward motion slows, but $D_1$ remains attached to the cylinder until the fifth frame. Vortex $D_1$ is shed at some point between the fifth and sixth frames. This vortex is relatively weak and appears to have been `prematurely' pulled away from the cylinder under the influence of its larger counterclockwise neighbor $C_2$. Then, in the remainder of the cylinder's upstroke, clockwise vorticity continues to develop along its top edge and remains connected to $D_1$ via a braid shear layer. Once again, the stronger vortex $D_2$ is shed not under the influence of a neighboring vortex, but due to the `jerk' of the cylinder, now in the opposite direction. This cylinder motion results in $D_2$ being a relatively strong counterclockwise vortex, analogous to the clockwise vortex $B_2$. As the cylinder executes its downward stroke, $D_2$ is once again positioned to draw away the counterclockwise vorticity, which can be seen developing in the ninth frame, labeled $E_1$, exactly analogous to $C_1$ in the first frame.

These observations show that counterclockwise vorticity is mostly generated during the downstroke, while clockwise vorticity is mostly generated during the upstroke. At each turning point, the `jerk' causes a strong vortex to be shed, which goes on to become the larger vortex in each pair comprising the 2P vortex street. A corresponding smaller vortex is shed during roughly the first half of each stroke, and this weak vortex rapidly convects away from the cylinder under the influence of the stronger neighbor shed just previously.

\subsection{P+S mode}
\begin{figure}
\centering
  \includegraphics[scale=0.9]{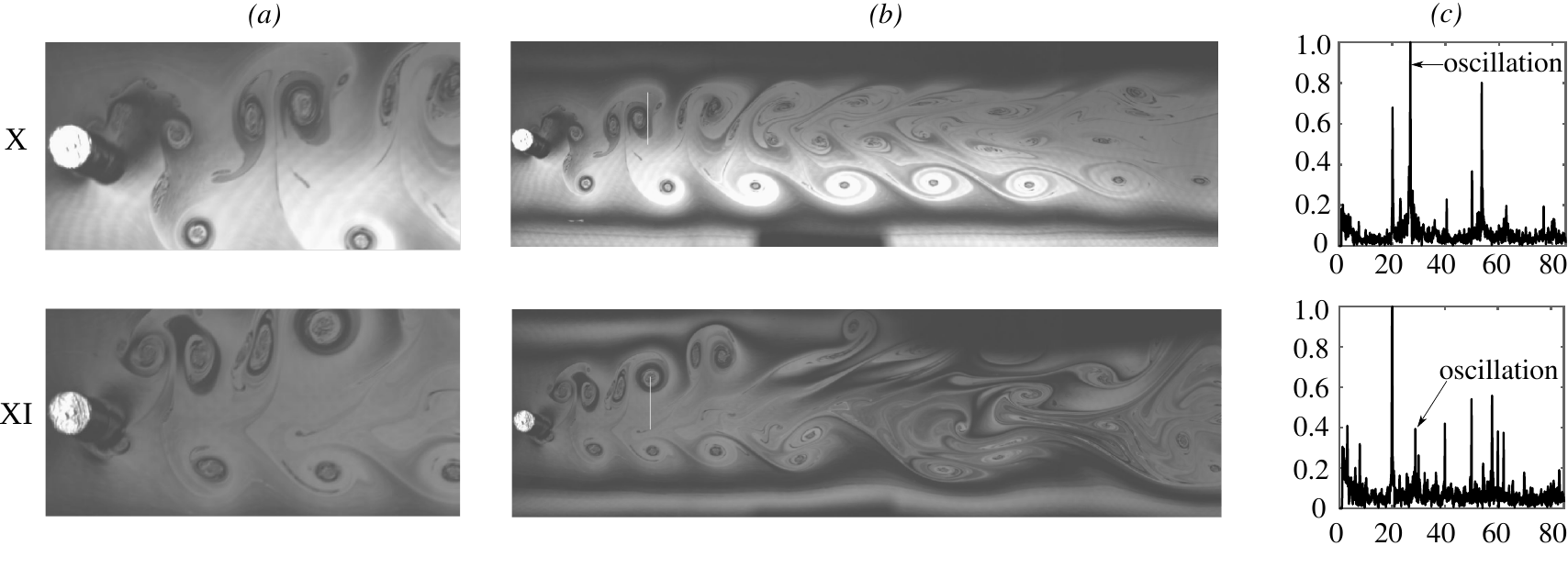}
  \caption{Representative cases of P+S vortex wakes.  Rows are labeled according to case with parameters X: $(f^*, A^*) \approx (1.28, 0.55)$ at $t^* = 0.862$ and XI: $(f^*, A^*) \approx (1.53, 0.63)$ at $t^* = 0.180$; cases are identified with the same labels in figure~\ref{fig:map}.  Columns show (a)~the near- and mid-wake regions (image size: 600 \by 300 pixels $\approx 10D \by 5D$), (b)~the extended wake region (image size: 2053 \by 600 pixels $\approx 34.2D \by 10D$), and (c) the normalized power spectrum for each case.  Contrast has been enhanced in the wake images to improve clarity. }
\label{fig:P+S}
\end{figure}

A 
`P+S' mode, in which three vortices are shed per oscillation cycle, was observed at high oscillation frequencies and moderately high amplitudes of oscillation. Two representative examples of P+S wakes are shown in figure \ref{fig:P+S}. The vortices that formed in the P+S mode were observed to be quite sensitive to perturbations and would often become significantly distorted in the mid- to far- wake; see, e.g., case~XI in figure~\ref{fig:P+S}. The near-wake regions show that the pair of vortices shed from one side of the cylinder often had `stray' concentrations of vorticity manifested as ``cat's eye'' shear layers or even small individual vortices. 
This additional vorticity tended to have a noticeable effect on wake periodicity and stability, as illustrated by the presence of multiple incommensurate peaks in the power spectrum for case~XI in figure~\ref{fig:P+S}(c).
In general, the P+S wakes observed in these experiments were not as regular as the 2P wakes 
and, in contrast to the 2S and 2P mode wakes, the power spectrum for case~XI is not dominated by the oscillation frequency.

\begin{figure}
\centering
  \includegraphics[scale=0.9]{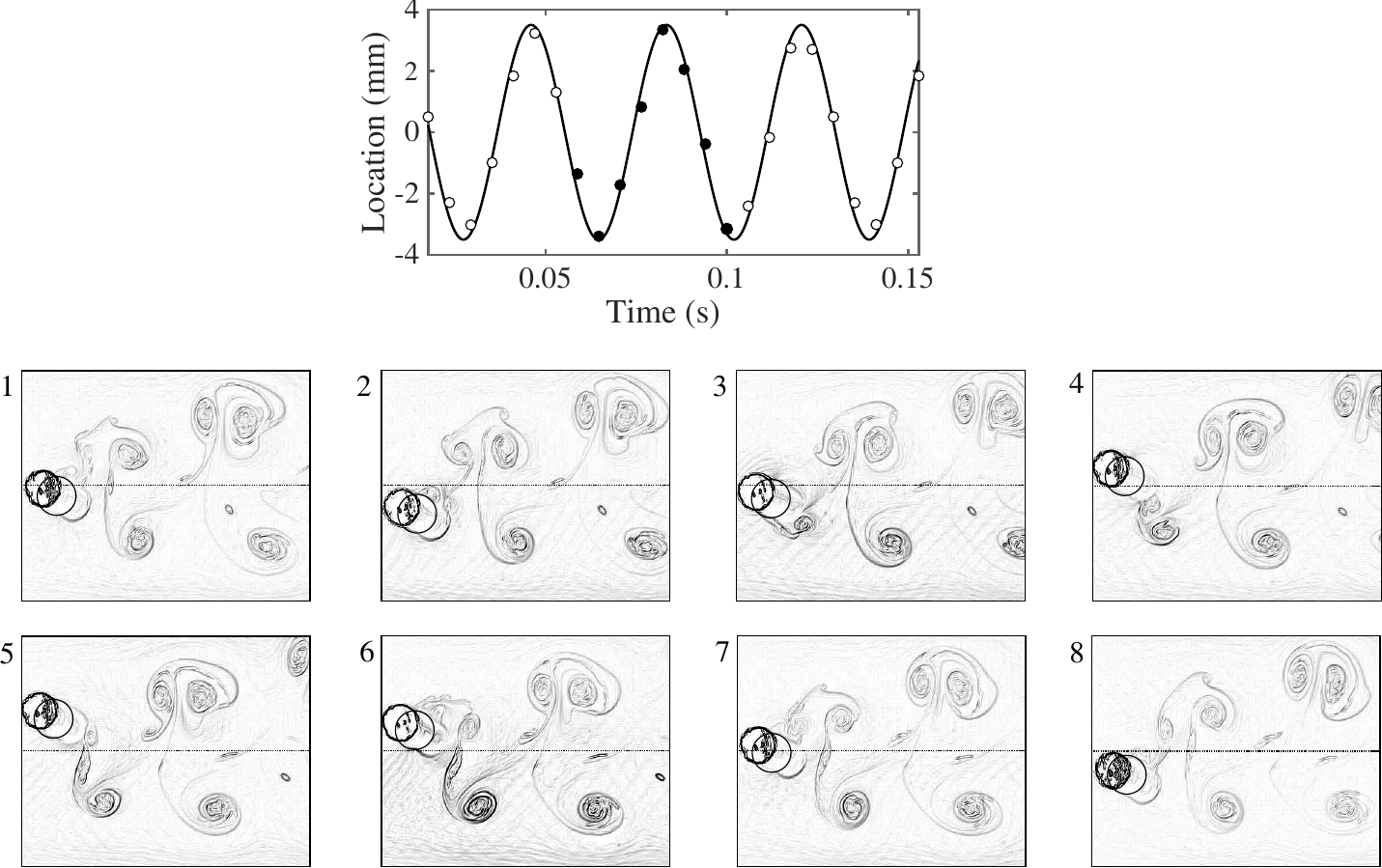}
  \caption{Vortex formation process in the `P+S' mode for case~X in figure~\ref{fig:P+S}. Top panel: Cylinder positions in time (solid circles) corresponding to the panel images; solid line shows \added[id=A]{a} sine curve fit to the \added[id=A]{entire 8.2-second} observed trajectory.  
  Numbered panels: flow structure for $t^*_i = \{ 0.107, 0.265, 0.423, 0.580, 0.738, 0.896, 1.054, 1.211\}$, with $i$ given by the panel label. Each wake image measures 599 \by 398 pixels $\approx 10D \by 6.6D$). Cylinder cross-section in contact with the soap film has been marked by a filled circle. Images have been processed to highlight flow structure. }
\label{fig:P+S_detail}
\end{figure}

The P+S wake formation process for case~X is illustrated in figure~\ref{fig:P+S_detail} over one period of cylinder motion.  The similarities between frames~1 and 8 clearly shows the (near) periodicity of the shedding process in this case and allows one to view frame~1 as a continuation of frame~7.   
For the sequence shown here, the first coherent vortex, with counterclockwise rotation, appears to have formed between times $t^*_2$ and $t^*_3$, either during the turning point of the downstroke or early in the upstroke.  A weak clockwise vortex was shed very quickly thereafter, between $t^*_3$ and $t^*_4$. A second, much stronger clockwise vortex began forming during the second half of the upstroke and was shed around $t^*_5$ at the turning point of the upstroke.  These two vortices merged into one clockwise vortex within one period, as can be seen in the progression from $t^*_7$ to $t^*_8$ or $t^*_1$ (by making use of the periodicity of the shedding process), and then to $t^*_2$.  The final counterclockwise vortex (of the period) was shed during the first half of the downstroke (see frames~6--7); the proximity of this vortex to the strong clockwise vortex led to these two vortices moving into the wake as a pair.    


As figure \ref{fig:map} shows, the P+S region of the parameter space manifested itself in our experiments as a narrow horizontal band in the high-frequency, mid-amplitude range of the $\faspace$ plane. This range of parameter space corresponds roughly to a portion of the coalescing P+S mode in \cite{Williamson1988} and a portion of the unlabeled region in \cite{Leontini2006}. The irregular behavior of the P+S mode in our system thus bears some resemblance to prior results.


We speculate that the P+S wake region of parameter space for an oscillating cylinder in a soap film extends into the high-frequency, high-amplitude region of the $\faspace$ plane 
that we were not able to explore. 
If this region of parameter space does indeed consist of P+S mode wakes, it would \deleted[id=R3]{the} increase the similarities between our results and those of \cite{Williamson1988} and \cite{Leontini2006}. 
It should be possible to modify the inclined soap film channel to access this region of parameter space. However, in the interest of keeping as many experimental variables constant as possible (e.g., angle of inclination, soap solution composition, diameter of the cylinder, approximate average speed of the soap solution, length and width of the test section) across as much of the $\faspace$ space as possible, we have not pursued such modifications in this work.

\subsection{2T mode}

\begin{figure}
\centering
  \includegraphics[scale=0.9]{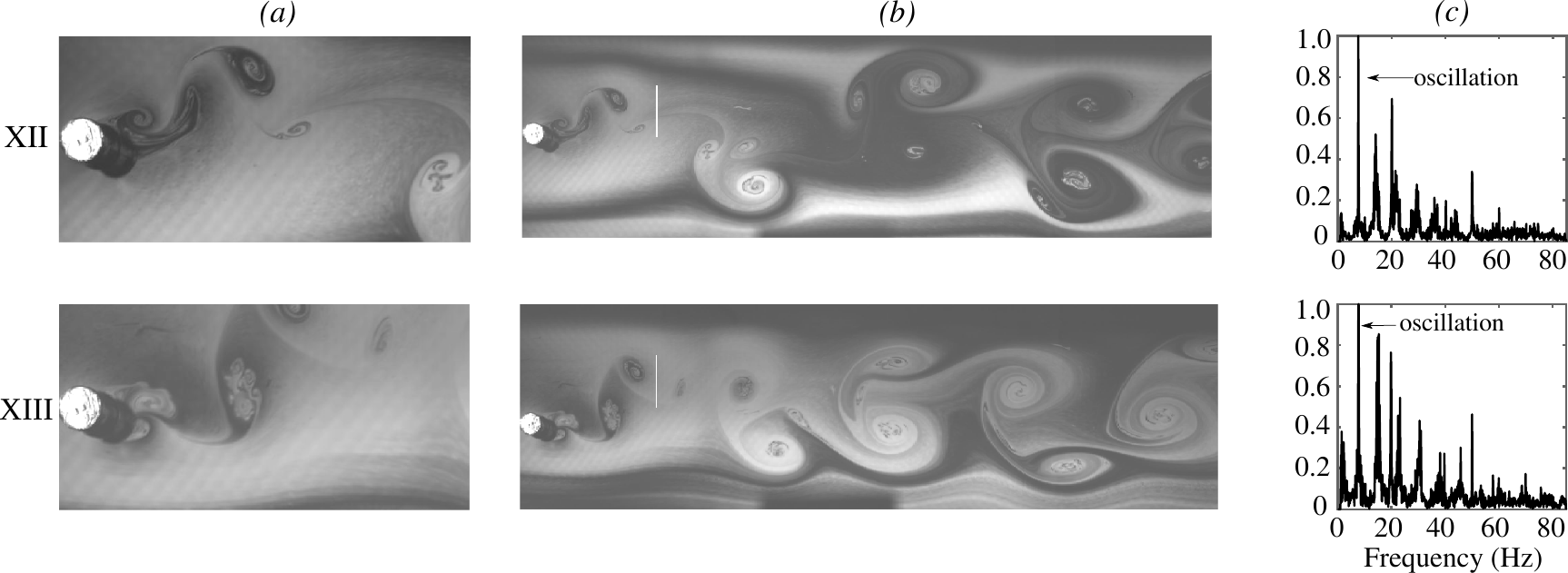}
  \caption{Representative cases of 2T vortex wakes.  Rows are labeled according to case with parameters XII: $(f^*, A^*) \approx (0.39, 0.79)$ at $t^* = 0.965$ and XIII: $(f^*, A^*) \approx (0.41, 1.10)$ at $t^* = 0.192$; cases are identified with the same labels in figure~\ref{fig:map}.  Columns show (a)~the near- and mid-wake regions (image size: 600 \by 300 pixels $\approx 10D \by 5D$), (b)~the extended wake region (image size: 2053 \by 600 pixels $\approx 34.2D \by 10D$), and (c) the normalized power spectrum for each case.  Contrast has been enhanced in the wake images to improve clarity. }
  \label{fig:2T}
\end{figure}

Another synchronized 
pattern of vortex-shedding was observed 
for the region of parameter space labeled `2T' in figure~\ref{fig:map},
in which six vortices were shed during each cycle of the cylinder's oscillation. As shown for two examples in figure~\ref{fig:2T}, these six vortices organized into two groups of three vortices each.  As the vortices moved into the far wake region their structure became noticeably distorted, yet they still maintained a clear `2T' pattern consisting of two sets of vortex triplets.  
The dominant frequency in the power spectra, shown in figure~\ref{fig:2T}(c), was again the oscillation frequency of the cylinder, showing a wake structure that was synchronized with the cylinder motion. 

\begin{figure}
\centering
  \includegraphics[scale=0.9]{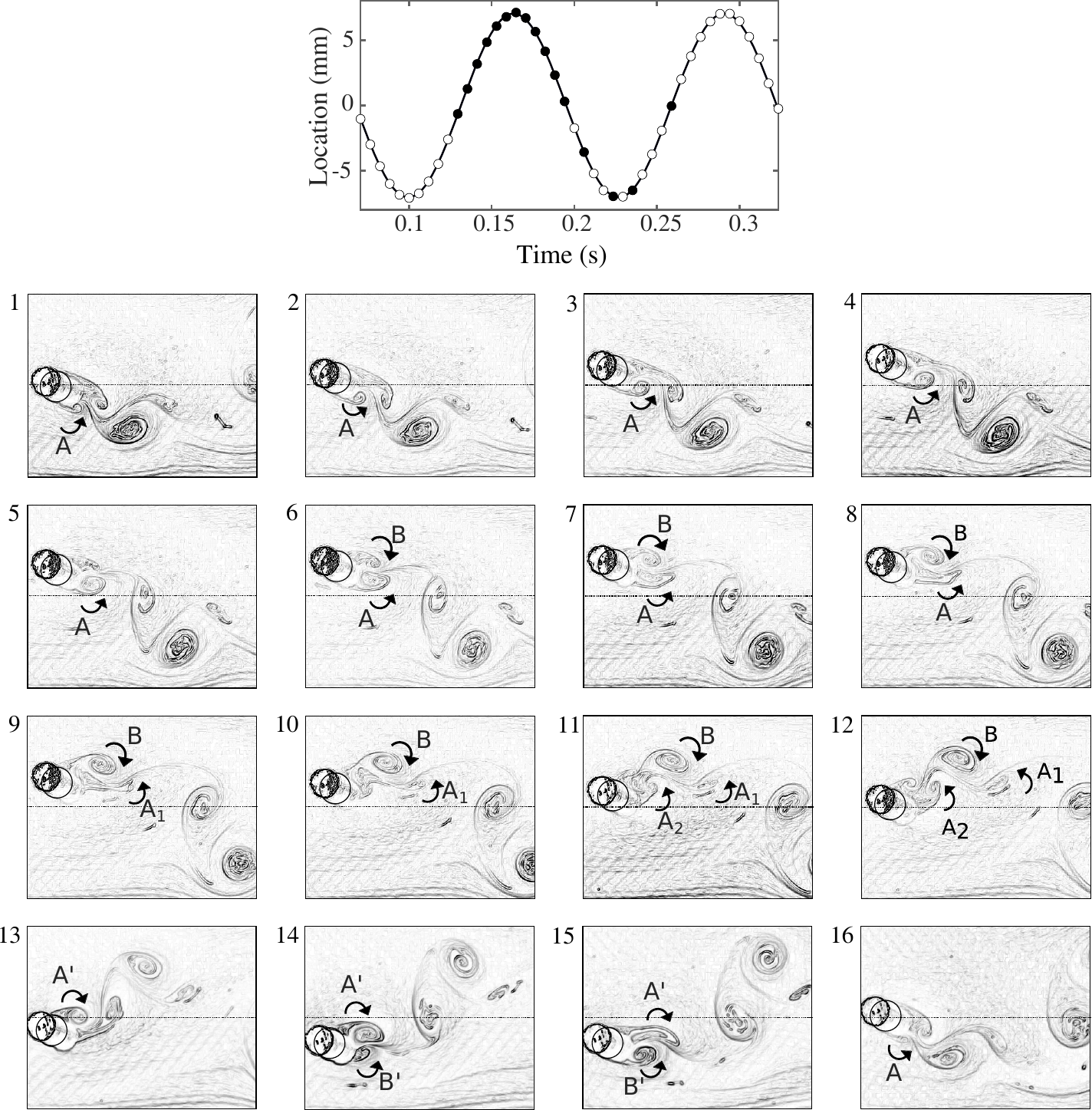}
  \caption{
Vortex formation process during \replaced[id=R3]{one }{a single half-}cycle of cylinder motion for the 2T wake case~XIII from figure~\ref{fig:2T}. 
\deleted[id=A]{The period of the cylinder's motion was 0.129 seconds; with a frame rate of 170 Hz, a half-cycle was captured in 11.0 frames, and 12 successive frames are shown here.}
  Top panel: Cylinder positions in time (solid circles) corresponding to the \replaced[id=A]{wake}{panel} images; solid line shows \added[id=A]{a} sine curve fit to the \added[id=A]{entire 8.2-second} observed trajectory.
  Numbered panels: flow structure for $t^*_i = \{ 0.477,0.522,0.568,0.613,0.659,0.704,0.749,0.795,0.840,0.885,0.931,0.977\added[id=R3]{, 1.067, 1.204, 1.294, 1.476}\}$, with $i$ given by the panel label. 
\added[id=R3]{Frames $i=1,...,12$ show successive images corresponding to a half-cycle. Frames $i=13,...,16$ were selected from the second half-cycle to highlight the  symmetry of this mode: compare panel 3 with 13, 6 with 14, 8 with 15, and 1 with 12 and 16. }
Each wake image measures \replaced[id=A]{500 \by 400 pixels}{599 \by 398 pixels} \replaced[id=A]{$\approx 8.25D \by 6.6D$}{$\approx 10D \by 6.6D$}). Cylinder cross-section in contact with the soap film has been marked by a filled circle. Images have been processed to highlight flow structure.  
 }

\label{fig:2Tdetail}
\end{figure}

\added[id=R3]{Panels numbered 1--12 in} figure \ref{fig:2Tdetail} show the formation of vortices during one half-cycle of the cylinder's motion for case~XIII, comprising 7 frames of the upstroke and 5 frames of the downstroke. In the first frame, the triplet formed in the previous half-cycle can be seen, and in frames 2--4 this triplet moves downstream away from the cylinder while the counter-clockwise vortex labeled~$A$ develops on the lower surface. As the cylinder decelerates to a complete stop and turns around in frames 5 through 8, $A$ remains attached while 
a clockwise vortex, labeled~$B$, rolls up on the upper surface of the cylinder and progressively strengthens.  By frame 8, when the cylinder has begun the downstroke, $B$ has become strong enough to distort $A$, which in frame~9 has been pinched off into a weak counter-clockwise vortex labeled $A_1$ that rapidly ejects away from the cylinder. As the cylinder accelerates downward, $B$ slowly detaches from the cylinder while the attached remainder of $A$ rolls up into another counter-clockwise vortex, labeled~$A_2$. By frame~12, the triplet comprised of $\{A_1, B, A_2 \}$ can be clearly seen\deleted[id=R3]{, and a mirror image of the same process begins to occur in the next half-cycle of cylinder motion}. 

\added[id=R3]{Panels labeled 13--16 in figure~\ref{fig:2Tdetail} were selected from the remaining half-cycle of the cylinder's motion  to show the glide-reflective symmetry of this mode. Frame 13 is approximately a `mirror image' of frame 3, frame 14 of frame 6, and frame 15 of frame 8;  frame 16 is nearly identical to frame 1. The vortices in frame 13--15 have been labeled with a prime to indicate that they are analogous to the previously-seen vortices (albeit with opposite sign). For example, between frames 14 and 15, clockwise vortex $A'$ begins to stretch and split into two, similar to the way counterclockwise vortex $A$ stretches and splits into two between frames 6 and 8.}

This vortex formation process for the 2T wakes appears to be quite regular, and it is our observation that the irregularity of the far wake in the 2T region --- compare, e.g., figure~\ref{fig:2T} with figure~\ref{fig:strong2P} --- was caused not by aperiodicity in the vortex shedding but by sensitivity of the subsequent vortex arrangements to disturbances in the flow.

This type of high-amplitude synchronized mode, in which a triplet of vortices forms from the vorticity shed during each half-cycle of cylinder motion, was first observed in experiments with a circular cylinder that was flexibly mounted and allowed to oscillate in both the streamwise and transverse directions \citep{Williamson2004a}. The same mode of vortex-shedding has been observed in three-dimensional computations of the flow over a cylinder undergoing controlled `figure-8' motion oscillations including both streamwise and transverse oscillation \citep{Zhi-yong2010}. Our results show that in-line movement of the cylinder is not necessary for the appearance of the 2T mode, at least for this soap film system. \cite{Leontini2006} did not include in their study the region of  parameter space for which we observed the 2T mode ($f^* < 0.5$); it is possible that the 2T mode would be found in low-Reynolds number two-dimensional Navier-Stokes computations for $f \approx \fst/3$.

\subsection{Transitional (tr) mode}
\begin{figure}
\centering
  \includegraphics[scale=0.9]{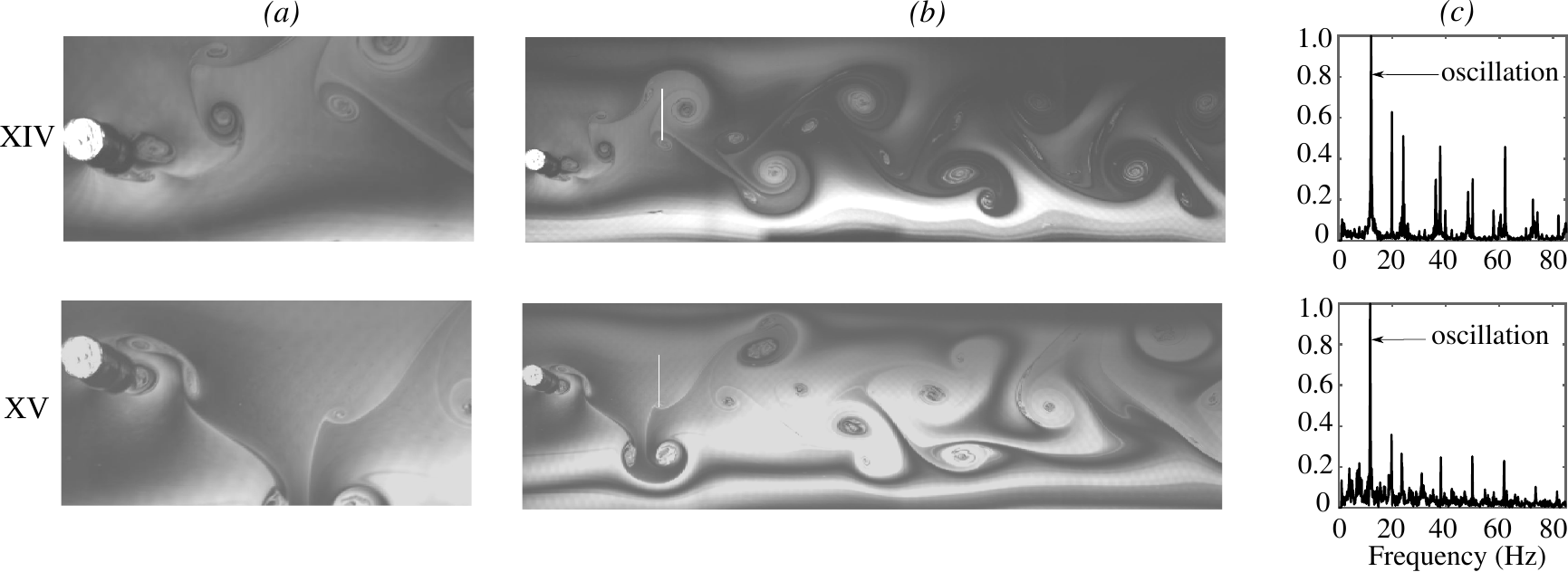}
  \caption{Representative cases of transitional (tr) vortex wakes.  Rows are labeled according to case with parameters XIV: $(f^*, A^*) \approx (0.64, 1.26)$ at $t^* = 0.264$ and XV: $(f^*, A^*) \approx (0.60, 0.94)$ at $t^* = 0.709$; cases are identified with the same labels in figure~\ref{fig:map}.  Columns show (a)~the near- and mid-wake regions (image size: 600 \by 300 pixels $\approx 10D \by 5D$), (b)~the extended wake region (image size: 2053 \by 600 pixels $\approx 34.2D \by 10D$), and (c) the normalized power spectrum for each case.  Contrast has been enhanced in the wake images to improve clarity. }
\label{fig:transitional}
\end{figure}

A portion of the primary synchronization region is marked `transitional' in figure~\ref{fig:map}, representing a transition from the synchronized 2T mode to the synchronized 2P mode as the oscillation frequency was increased. This transition regime consisted of synchronous vortex shedding \replaced[id=A]{with complex and often}{ but} asynchronous wake structure\deleted[id=A]{, in which no single repeatable wake pattern was consistently identified}. Two examples of transitional wakes are shown in figure~\ref{fig:transitional}. The power spectra in the transitional regime continued to be dominated by the cylinder oscillation frequency, leading us to classify these wakes as `synchronized' despite the \added[id=A]{typical} lack of coherence in the resulting wake structure.

In some cases, intermittent loss of lock-on was observed during the course of a single (8.2-seconds-long) experiment: for short periods of time, vortex shedding would occur clearly synchronized with the cylinder's motion, but this lock-on would be lost after irregular periods, leading to an overall incoherent wake structure. Case~XV in figure \ref{fig:transitional} is an example of such a result: the near wake contains a clear vortex pair shed in the most recent cycle, but in the mid-wake there are several haphazardly-arranged vortices from the previous few cycles. This example shows that, although vortex shedding did occur in correlation with the cylinder's motion, it was  not necessarily periodic, with different numbers of vortices being shed during various cycles of the cylinder's motion.

In other cases, we observed what appear to be other modes of synchronization, such as periodic vortex shedding that does not fit the descriptions of modes 2S, P+S, 2P or 2T; see,  e.g., case~XIV in figure~\ref{fig:transitional}. However, these modes were either temporally short-lived --- i.e., consistent wake structure was not sustained for an extended period of time --- or they did not persist over a large enough range of the frequency-amplitude space to be properly identified with our limited resolution in $\faspace$ space. 

It is likely that a closer examination of the transitional region with small increments in oscillation frequency will show clearly defined synchronization regions corresponding to commensurate ratios of $f^*$. \cite{Schnipper2009} found periodically repeating vortex patterns in the wake of an oscillating airfoil at low Reynolds numbers with up to 16 vortices per period. Such high-order synchronization modes have yet to be observed in the wakes of transversely-oscillating cylinders.


\subsection{Coalescing (clsc) mode}
\begin{figure}
\centering
  \includegraphics[scale=0.9]{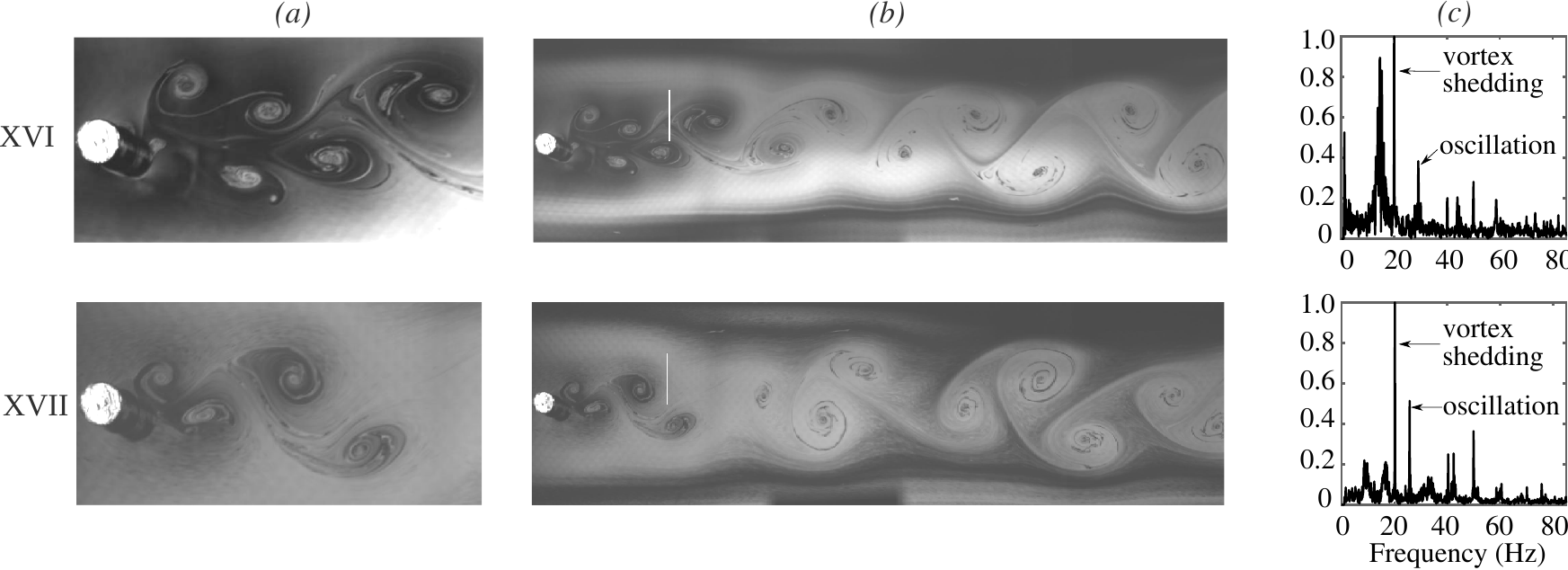}
  \caption{Representative cases of coalescing vortex wakes.  Rows are labeled according to case with parameters XVI: $(f^*, A^*) \approx (1.50, 0.39)$ at $t^* = 0.196$ and XVII: $(f^*, A^*) \approx (1.36, 0.24)$ at $t^* = 0.935$; cases are identified with the same labels in figure~\ref{fig:map}.  Columns show (a)~the near- and mid-wake regions (image size: 600 \by 300 pixels $\approx 10D \by 5D$), (b)~the extended wake region (image size: 2053 \by 600 pixels $\approx 34.2D \by 10D$), and (c) the normalized power spectrum for each case.  Contrast has been enhanced in the wake images to improve clarity. }
\label{fig:coalescing}
\end{figure}

\begin{figure}
\centering
  \includegraphics[scale=0.9]{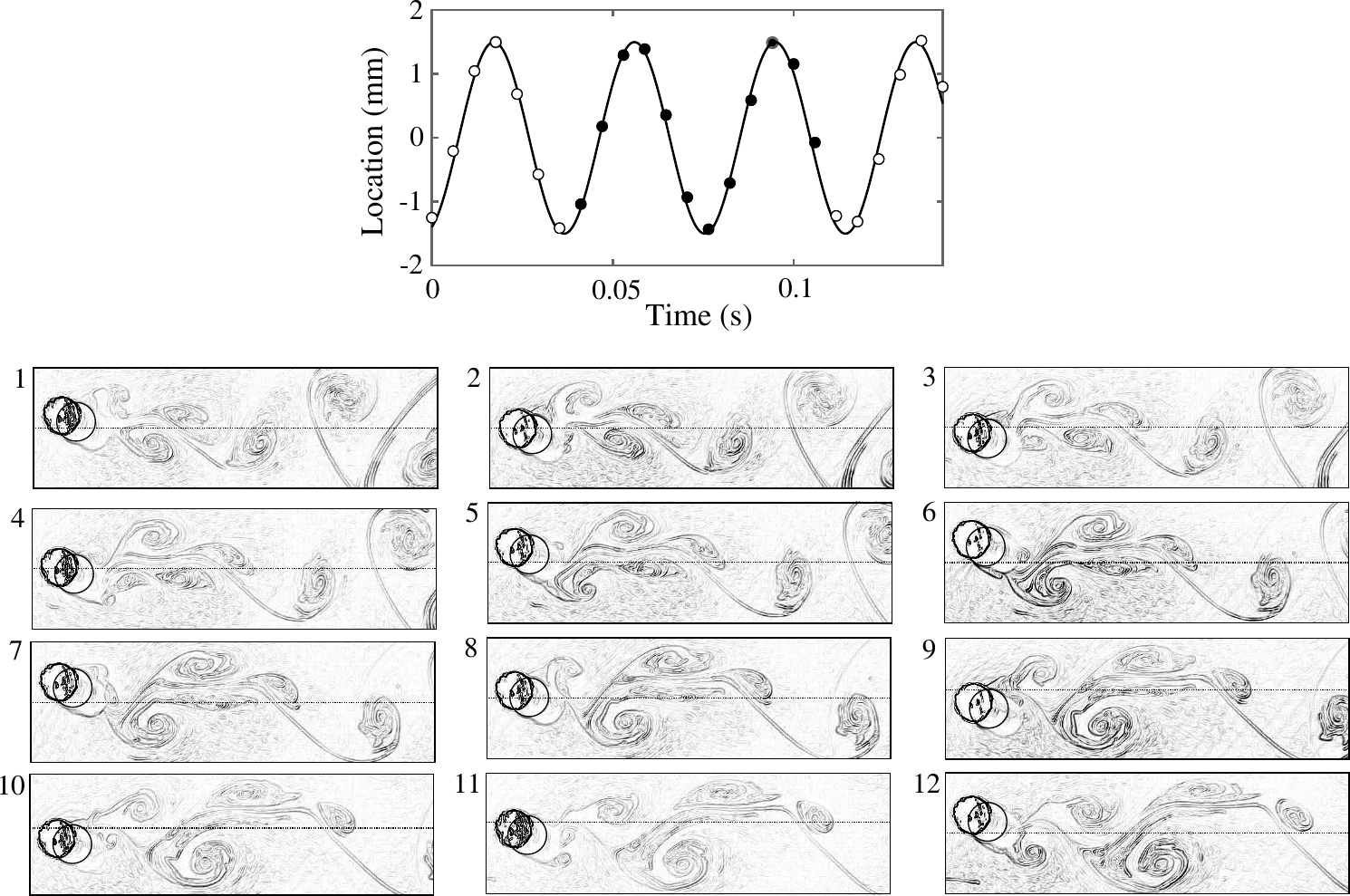}
  \caption{Vortex formation process in the `coalescing' mode for case~XVII in figure~\ref{fig:coalescing}. 
  Top panel: Cylinder positions in time (solid circles) corresponding to the panel images; solid line shows \added[id=A]{a} sine curve fit to the \added[id=A]{entire 8.2-second} observed trajectory.
  Numbered panels: flow structure for $t^*_i = \{ -0.032, 0.120, 0.271, 0.422, 0.574, 0.725, 0.876, 1.028, 1.180, 1.330, 1.482, 1.633\}$, with $i$ given by the panel label. Each wake image measures 599 \by 398 pixels $\approx 10D \by 6.6D$). Cylinder cross-section in contact with the soap film has been marked by a filled circle. Images have been processed to highlight flow structure.
 }
\label{fig:coalescingdetail}
\end{figure}

The low-amplitude ($A^*<0.5)$, high-frequency ($f^* > 1$) region of figure \ref{fig:map} 
was dominated by a `coalescing' mode of vortex shedding.  As illustrated for two example cases in figure~\ref{fig:coalescing}, this regime is characterized by multiple vortices being shed in each oscillation cycle 
with no discernible consistency in the near-wake patterns.  In contrast to every regime considered thus far, the oscillation frequency does not dominate the power spectra, showing that these wakes are not synchronized with the cylinder motion.  

A representative temporal sequence over approximately 1.5 oscillation cycles of case~XVII is shown in figure~\ref{fig:coalescingdetail}.  The lack of synchronization between the cylinder motion and the initial wake structure 
can be seen by comparing panels~3 and 10, which occur at approximately the same relative time in the cylinder's motion ($t^*_3=0.271$ and $t^*_{10}=1.330$, approximately one full period later).   
After some complicated inter-vortex motions in the near-wake, the vortices then coalesced together into a `2S-like' arrangement, with multiple individual vortices combining to form each element of the newly-formed vortex street. 
This amalgamation of many small vortices into large vortices was also reported by \cite{Williamson1988} in the low-amplitude, high-frequency region of the parameter space. The coalescence of small vortices into a new `vortex street'-like arrangement downstream may occur due to the same processes that lead to the well-known breakdown of the von~\karman\ street and its subsequent reorganization into a `secondary vortex street' (see, for example, \cite{Kumar2012}). 

The coalescing mode of vortex shedding exhibits both disorder --- in the rapid shedding of multiple vortices into the near wake in each oscillation cycle --- and order, in the later reorganization of those vortices into a (relatively) regular vortex street. A von~\karman-like wake was the most commonly observed structure for the coalesced wakes, but the structure was also observed to vary with time.  In case~XVI, for example, the merged vortices formed a very consistent 2S-like far wake.  Despite this coalesced structure not emerging until downstream of the interrogation window, the power spectrum for case~XVI in figure~\ref{fig:coalescing}(c) shows a strong peak at the frequency of the far wake, which is lower than both $f$ and $\fst$.  For case~XVII, in contrast, the merged vortices cluster into various patterns that vary with time; the instant shown in figure~\ref{fig:coalescing}(b) suggests the vortices were arranging into a pattern that is more `2T-like' than `2S-like'.  This lack of a consistent wake pattern is reflected in the power spectra being dominated by $\fst$ and $f$ for case~XVII.

\subsection{Perturbed von \karman\ (pvK) mode}

\begin{figure}
\centering
  \includegraphics[scale=0.9]{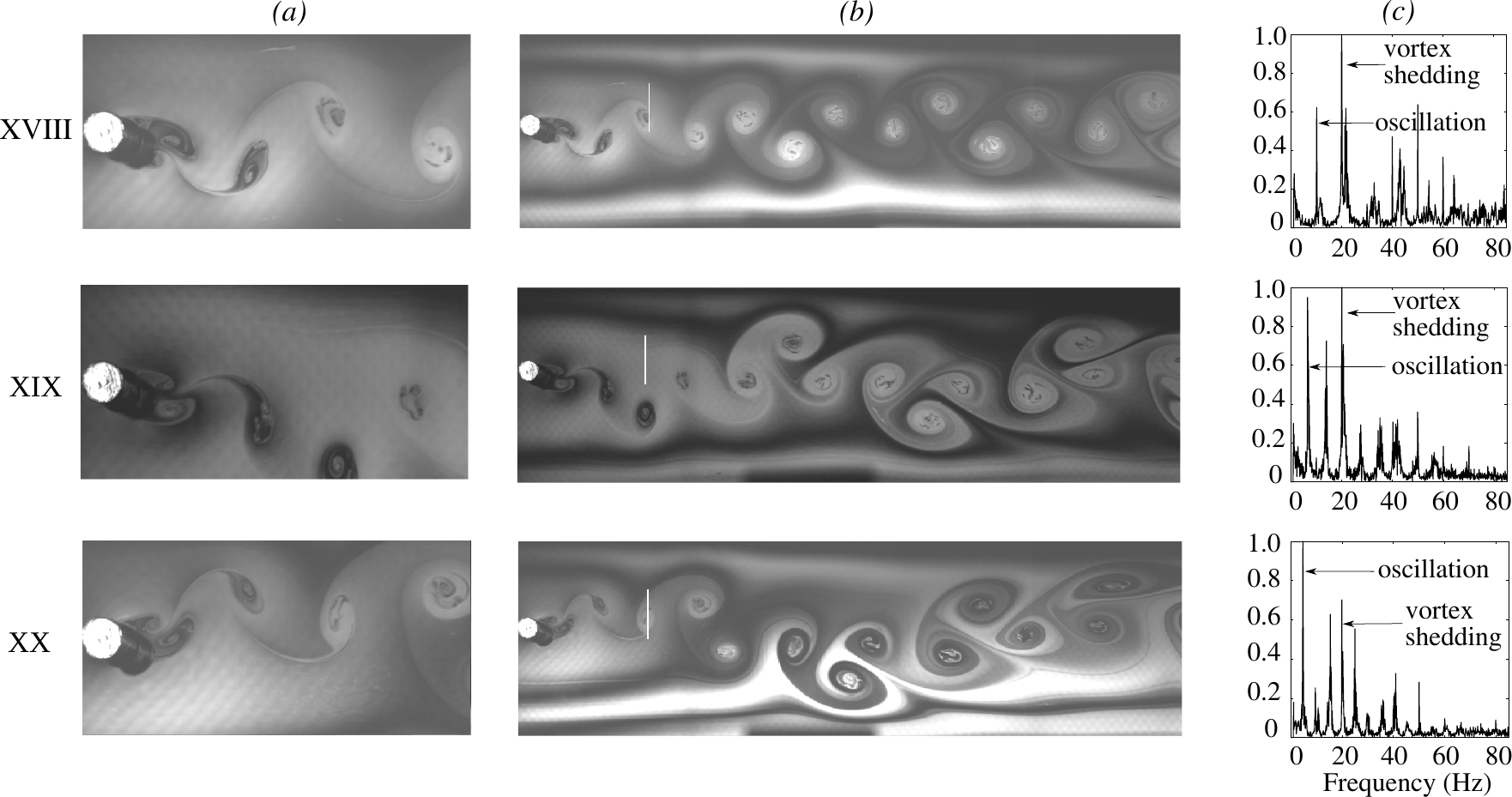}
  \caption{Representative cases of perturbed von \karman\ vortex wakes.  Rows are labeled according to case with parameters XVIII: $(f^*, A^*) \approx (0.48, 0.16)$ at $t^* = 0.471$, XIX: $(f^*, A^*) \approx (0.32, 0.55)$ at $t^* = 0.521$, and XX: $(f^*, A^*) \approx (0.24, 0.94)$ at $t^* = 0.943$; cases are identified with the same labels in figure~\ref{fig:map}.  Columns show (a)~the near- and mid-wake regions (image size: 600 \by 300 pixels $\approx 10D \by 5D$), (b)~the extended wake region (image size: 2053 \by 600 pixels $\approx 34.2D \by 10D$), and (c) the normalized power spectrum for each case.  Contrast has been enhanced in the wake images to improve clarity. }
\label{fig:pvK}
\end{figure}

At extremely low oscillation frequencies, the wake does not synchronize with the motion of the cylinder; the transverse motion is too slow to affect the vortex shedding, which remains independent of the cylinder's motion. Instead, vortex shedding continues to occur at its natural frequency and does not synchronize with the cylinder's motion. As shown by the three examples in figure~\ref{fig:pvK}, the wake structure resembles that of a fixed cylinder but superimposed along a sinusoidal `centerline' caused by the cylinder's motion. 
We thus refer to these wakes as being in a `perturbed von~\karman' (pvK) mode.  
The organization of these vortices into distinct patterns in the far wake is the result of post-shedding vortex dynamics, rather than the interplay between cylinder motion and vortex shedding, as was the case with the synchronized modes discussed earlier. In the pvK  region of parameter space labeled in figure~\ref{fig:map}, the peak in the power spectrum associated with the Strouhal frequency ($\fst\approx 20$\,Hz) is comparable to the peak occurring at the oscillation frequency, as can be seen in figure \ref{fig:pvK}(c).

Although the perturbed von \karman\ mode can look similar to the 2S mode at low amplitudes\added[id=R2]{,} \deleted[id=R2]{--- compare case~XVIII from figure~\ref{fig:pvK} with case~I from figure \ref{fig:2S} ---} there is a crucial difference between these modes. In the 2S mode, clockwise vortices are only shed during the upstroke and counterclockwise ones only during the downstroke; in the perturbed von~\karman\ mode, vortices of either sign are shed during any part of the cylinder's cycle.  
\added[id=R2]{At these low amplitudes, the power spectra of wakes in the (unsynchronized) pvK mode exhibit stronger peaks at the shedding frequency than at the oscillation frequency (see figure \ref{fig:pvK}(c) panels XVIII and XIX), in contradistinction to wakes in the synchronized modes, whose spectra always have a stronger peak at the oscillation frequency. In particular, this criterion was used to classify wakes as pvK as opposed to 2S in cases where the two are visually similar (compare figure~\ref{fig:pvK}~XVIII~(b) with figure~\ref{fig:2S}~I~(b)). }


At high amplitudes, the window in which the perturbed von \karman\ mode occurs shifts further toward the low-frequency edge of the parameter space, but it still exists; cases~XVII and XVIII represent experiments at $A^* = 0.55$ and $0.94$, respectively. The resulting wakes shown in figure \ref{fig:pvK} can be interpreted as long-wavelength, high-amplitude sinusoidal perturbations of a double row of counter-rotating vortices. \added[id=R2]{At these high amplitudes in the pvK mode, the peak at the shedding frequency is not necessarily stronger than that at the oscillation frequency, but the wake is visually identifiable as a `sinusoidally perturbed' von \karman\ street in which the vortex shedding occurs at $\fst$ rather than at the oscillation frequency.}

The infinite counter-rotating double row of point vortices in an ideal fluid is known to be unstable to large-amplitude sinusoidal disturbances. However, the growth rate of disturbances decreases with increasing wavelength \citep{Aref1995}, which implies that the \karman\ street is least unstable to the longest-wavelength disturbances. Therefore, we believe that the persistence and repeatability of the perturbed von~\karman\ mode is due to the low growth rate of sinusoidal disturbances when the wavelength is large relative to the distance between successive vortices. At these low growth rates, the time scale of the break-up of the vortex street is much longer than the convective time scale, allowing the vortices to be swept downstream long before the instability can set in.

\section{Conclusions}
\label{sec:conclusions}

We have presented results for the wake structure of a circular cylinder undergoing controlled transverse oscillations with normalized amplitudes in the range $0.1<A/D<1.3$ and normalized frequencies in the range $0.2<f/f_{St}<1.8$ at $Re = 235 \pm 14$ in a flowing soap film. 
This Reynolds number value complements previous experimental results \citep{Williamson1988} and facilitates comparisons with computational results \citep{Leontini2006}.  An unconstrained wake is likely to develop three-dimensional features at this value of Re \citep{Williamson1996a, Leweke1998}, but at $\text{Re} = 235$ a two-dimensional analysis remains physically relevant \citep{Leontini2006}.  
The ranges for $f^*$ and $A^*$ were chosen to focus on the primary synchronization region, which is of significant interest owing to its connection with the phenomenon of vortex-induced vibrations in freely-oscillating bluff bodies.

Our results show \deleted[id=A]{that} the wake of a circular cylinder \replaced[id=A]{in a soap film being}{ becomes} `locked-on' to the cylinder's motion over a significant portion of the explored $\faspace$ space. Lock-in leads to a variety of synchronized patterns of coherent vortices in the wake.
For $A^*\lesssim 0.5$ and $0.6 \lesssim f^*\lesssim 1.1$, the wakes synchronized to a 2S mode, with a wake frequency corresponding to the cylinder oscillation, not the vortex shedding frequency for the corresponding fixed cylinder (i.e., not $\fst$).  Increasing the amplitude to $A^*\gtrsim 0.5$ with $f^* \approx 1$ led to the production of 2P wakes.  For the lower amplitudes and frequencies in this range, the secondary vortex in each pair was quite weak (in the `weak 2P' region), and the boundary between the 2S and weak 2P structures is \added[id=R2]{highly} subjective. The remainder of the 2P regime gave a clear two-pair structure per shedding cycle, and the relative vortex motion gave evidence of both `orbiting' and `exchanging' modes.  With $A^*\gtrsim0.5$, increasing the oscillation frequency from the 2P regime produced P+S mode wakes.  Experimental limitations confined our observations of P+S wakes to $f^*\gtrsim 1.2$ and $0.5 \lesssim A^* \lesssim 0.7$; determining if P+S wakes can be produced in a soap film for $A^*\gtrsim 0.7$ remains to be determined.  Maintaining $A^*\gtrsim0.5$ and decreasing the oscillation frequency from the 2P regime led to the production of transitional wakes, for which vortex generation was synchronized to the cylinder oscillation but \replaced[id=A]{consistent wake patterns were not identified}{ the resulting wakes did not consist of clearly repeatable patterns}.  Decreasing the oscillation frequency further to $f^* \approx 1/3$ produced synchronized 2T wakes; this observation appears to be the first documentation of 2T wake production by a cylinder undergoing purely transverse oscillations.  

Moving sufficiently far from $f^*\approx 1$ produced wakes that were not synchronized with the cylinder motion.  For very low-frequency oscillations, vortices were shed at a frequency corresponding to that for a fixed cylinder, giving rise to perturbed von~\karman\ (pvK) wakes consisting of a 2S pattern superimposed on a sine wave.  For large oscillation frequencies ($f^*\gtrsim 1.1$) with $A^*\lesssim0.5$, each oscillation cycle produced multiple closely-spaced vortices that merged into a smaller number of large vortices, forming a coalesced (clsc) structure in the far wake that typically resembled a 2S wake.  
Even though the vortex shedding was not synchronized with the cylinder motion, both the pvK and clsc wakes typically displayed a stable, periodic structure in the far wake; in fact, these unsynchronized wakes tended to exhibit a more regular far-wake structure than the synchronized transitional wakes.  
For the unsychronized wakes, both the vortex shedding frequency, $\fst$, and the cylinder oscillation frequency, $f$, play an important role in the exact structure of the far wake; the details of this dependence remain to be determined.  

\added[id=A]{The results of our soap film experiments show strong correlations with prior experimental and computational results.  At the same time, there are several differences in these results that suggest interesting topics for further investigation.  The occurrence of the 2T mode wakes is one such topic.  }
\added[id=R1]{For another, while the shape of the synchronization region is consistent with previous results (compare the boundaries of the 2S and weak 2P regions in figure~\ref{fig:map} with those in figure~\ref{fig:WR-L}(a,b)), the synchronization region in the soap film covers a considerably wider range in $f^*$ than in the traditional flows.}
\added[id=A]{Finally, d}espite the physical constraints of the soap film system, which prevent any span-wise structure in the wake\added[id=R2]{ wider than the thickness of the film}, the results presented here are more similar to \deleted[id=R1]{the} three-dimensional experimental results \replaced[id=R1]{\citep{Williamson1988}}{ from \citet{Williamson1988}} than to \deleted[id=R1]{the} two-dimensional computational results \replaced[id=R1]{\citep{Leontini2006}}{ from \cite{Leontini2006}}\added[id=R2]{, which show at most a weak 2P mode \citep{Placzek2009, Dorogi2020}}.  An exploration of this intriguing dichotomy will be the subject of a future manuscript.  One major conclusion that can be drawn from this work is that the presence of \added[id=A]{span-wise} three-dimensional effects in the wake, \replaced[id=A]{such as the}{ in the form of} mode~A and mode~B instabilities \citep{Williamson1996a, Leweke1998}, \replaced[id=A]{is}{ are} not necessary for generation of a 2P wake structure.

\section*{Acknowledgements}
This material is based upon work supported in part by the National Science Foundation (NSF)'s Graduate Research Fellowship under Grant No. 1840995, NSF Grant No. DGE-0966125,  and the Virginia Tech Graduate Research Development Fund. The authors thank Dr\added[id=A]{s}.~Yanyun Chen\added[id=A]{, Saikat Basu, and Sunny Jung} for \replaced[id=A]{their}{ her} assistance in developing the experimental system.


\bibliographystyle{cas-model2-names}

\bibliography{OscillatingCylinderManuscript}

\begin{thebibliography}{51}
\expandafter\ifx\csname natexlab\endcsname\relax\def\natexlab#1{#1}\fi
\providecommand{\url}[1]{\texttt{#1}}
\providecommand{\href}[2]{#2}
\providecommand{\path}[1]{#1}
\providecommand{\DOIprefix}{doi:}
\providecommand{\ArXivprefix}{arXiv:}
\providecommand{\URLprefix}{URL: }
\providecommand{\Pubmedprefix}{pmid:}
\providecommand{\doi}[1]{\href{http://dx.doi.org/#1}{\path{#1}}}
\providecommand{\Pubmed}[1]{\href{pmid:#1}{\path{#1}}}
\providecommand{\bibinfo}[2]{#2}
\ifx\xfnm\relax \def\xfnm[#1]{\unskip,\space#1}\fi
\bibitem[{Antoine et~al.(2016)Antoine, de~Langre and Michelin}]{Antoine2016}
\bibinfo{author}{Antoine, G.O.}, \bibinfo{author}{de~Langre, E.},
  \bibinfo{author}{Michelin, S.}, \bibinfo{year}{2016}.
\newblock \bibinfo{title}{{Optimal energy harvesting from vortex-induced
  vibrations of cables}}.
\newblock \bibinfo{journal}{Proc. R. Soc. A Math. Phys. Eng. Sci.}
  \bibinfo{volume}{472}, \bibinfo{pages}{20160583}.
\newblock \URLprefix
  \url{https://royalsocietypublishing.org/doi/10.1098/rspa.2016.0583},
  \DOIprefix\doi{10.1098/rspa.2016.0583}.
\bibitem[{Aref(1995)}]{Aref1995}
\bibinfo{author}{Aref, H.}, \bibinfo{year}{1995}.
\newblock \bibinfo{title}{{On the equilibrium and stability of a row of point
  vortices}}.
\newblock \bibinfo{journal}{Journal of Fluid Mechanics} \bibinfo{volume}{290},
  \bibinfo{pages}{167}.
\newblock \URLprefix
  \url{http://www.journals.cambridge.org/abstract{\_}S002211209500245X},
  \DOIprefix\doi{10.1017/S002211209500245X}.
\bibitem[{Auliel et~al.(2015)Auliel, Castro, Sosa and Artana}]{Auliel2015}
\bibinfo{author}{Auliel, M.I.}, \bibinfo{author}{Castro, F.},
  \bibinfo{author}{Sosa, R.}, \bibinfo{author}{Artana, G.},
  \bibinfo{year}{2015}.
\newblock \bibinfo{title}{{Gravity-driven soap film dynamics in subcritical
  regimes}}.
\newblock \bibinfo{journal}{Physical Review E - Statistical, Nonlinear, and
  Soft Matter Physics} \bibinfo{volume}{92}, \bibinfo{pages}{1--8}.
\newblock \DOIprefix\doi{10.1103/PhysRevE.92.043009}.
\bibitem[{Basu and Stremler(2015)}]{Basu2015}
\bibinfo{author}{Basu, S.}, \bibinfo{author}{Stremler, M.A.},
  \bibinfo{year}{2015}.
\newblock \bibinfo{title}{{On the motion of two point vortex pairs with
  glide-reflective symmetry in a periodic strip}}.
\newblock \bibinfo{journal}{Physics of Fluids} \bibinfo{volume}{27}.
\newblock \DOIprefix\doi{10.1063/1.4932534}.
\bibitem[{Bishop and Hassan(1964)}]{Bishop1964}
\bibinfo{author}{Bishop, R.E.D.}, \bibinfo{author}{Hassan, A.Y.},
  \bibinfo{year}{1964}.
\newblock \bibinfo{title}{{The lift and drag forces on a circular cylinder
  oscillating in a flowing fluid}}.
\newblock \bibinfo{journal}{Proceedings of the Royal Society of London. Series
  A. Mathematical and Physical Sciences} \bibinfo{volume}{277},
  \bibinfo{pages}{51--75}.
\newblock \URLprefix
  \url{http://rspa.royalsocietypublishing.org/cgi/doi/10.1098/rspa.1964.0005},
  \DOIprefix\doi{10.1098/rspa.1964.0005}.
\bibitem[{Blackburn and Henderson(1999)}]{Blackburn1999}
\bibinfo{author}{Blackburn, H.M.}, \bibinfo{author}{Henderson, R.D.},
  \bibinfo{year}{1999}.
\newblock \bibinfo{title}{{A study of two-dimensional flow past an oscillating
  cylinder}}.
\newblock \bibinfo{journal}{J. Fluid Mech.} \bibinfo{volume}{385},
  \bibinfo{pages}{255--286}.
\newblock \DOIprefix\doi{10.1017/S0022112099004309}.
\bibitem[{Brika and Laneville(1993)}]{Brika1993}
\bibinfo{author}{Brika, D.}, \bibinfo{author}{Laneville, A.},
  \bibinfo{year}{1993}.
\newblock \bibinfo{title}{{Vortex-induced vibrations of a long flexible
  circular cylinder}}.
\newblock \bibinfo{journal}{Journal of Fluid Mechanics} \bibinfo{volume}{250},
  \bibinfo{pages}{481--508}.
\newblock \DOIprefix\doi{10.1017/S0022112093001533}.
\bibitem[{Couder and Basdevant(1986)}]{Couder1986}
\bibinfo{author}{Couder, Y.}, \bibinfo{author}{Basdevant, C.},
  \bibinfo{year}{1986}.
\newblock \bibinfo{title}{{Experimental and numerical study of vortex couples
  in two-dimensional flows}}.
\newblock \bibinfo{journal}{J. Fluid Mech.} \bibinfo{volume}{173},
  \bibinfo{pages}{225--251}.
\newblock \DOIprefix\doi{10.1017/S0022112086001155}.
\bibitem[{Couder et~al.(1984)Couder, Basdevant and Thome}]{Couder1984}
\bibinfo{author}{Couder, Y.}, \bibinfo{author}{Basdevant, C.},
  \bibinfo{author}{Thome, H.}, \bibinfo{year}{1984}.
\newblock \bibinfo{title}{{Sur l'apparition de couples solitaires de
  tourbillons dans les sillages bidimensionnels turbulents}}.
\newblock \bibinfo{journal}{Comptes Rendus l'Academie des Sci. Ser. II, Mec.
  Phys. Chim. Astron.} \bibinfo{volume}{299}, \bibinfo{pages}{1--15}.
\bibitem[{Couder et~al.(1989)Couder, Chomaz and Rabaud}]{Couder1989}
\bibinfo{author}{Couder, Y.}, \bibinfo{author}{Chomaz, J.},
  \bibinfo{author}{Rabaud, M.}, \bibinfo{year}{1989}.
\newblock \bibinfo{title}{{On the hydrodynamics of soap films}}.
\newblock \bibinfo{journal}{Phys. D Nonlinear Phenom.} \bibinfo{volume}{37},
  \bibinfo{pages}{384--405}.
\newblock \DOIprefix\doi{10.1098/rspl.1877.0052}.
\bibitem[{Dorogi and Baranyi(2020)}]{Dorogi2020}
\bibinfo{author}{Dorogi, D.}, \bibinfo{author}{Baranyi, L.},
  \bibinfo{year}{2020}.
\newblock \bibinfo{title}{{Identification of upper branch for vortex-induced
  vibration of a circular cylinder at Re=300}}.
\newblock \bibinfo{journal}{J. Fluids Struct.} \bibinfo{volume}{98},
  \bibinfo{pages}{103135}.
\newblock \URLprefix \url{https://doi.org/10.1016/j.jfluidstructs.2020.103135},
  \DOIprefix\doi{10.1016/j.jfluidstructs.2020.103135}.
\bibitem[{Eshraghi et~al.(2021)Eshraghi, Rajendran, Yang, Stremler and
  Vlachos}]{Eshraghi2021}
\bibinfo{author}{Eshraghi, J.}, \bibinfo{author}{Rajendran, L.K.},
  \bibinfo{author}{Yang, W.}, \bibinfo{author}{Stremler, M.A.},
  \bibinfo{author}{Vlachos, P.V.}, \bibinfo{year}{2021}.
\newblock \bibinfo{title}{On flowing soap films as experimental models of {2D}
  {N}avier-{S}tokes flows}.
\newblock \bibinfo{journal}{Experiments in Fluids} \bibinfo{note}{(in
  revision)}.
\bibitem[{Georgiev and Vorobieff(2002)}]{Georgiev2002}
\bibinfo{author}{Georgiev, D.}, \bibinfo{author}{Vorobieff, P.},
  \bibinfo{year}{2002}.
\newblock \bibinfo{title}{{The slowest soap-film tunnel in the Southwest}}.
\newblock \bibinfo{journal}{Review of Scientific Instruments}
  \bibinfo{volume}{73}, \bibinfo{pages}{1177--1184}.
\newblock \URLprefix \url{http://aip.scitation.org/doi/10.1063/1.1446040},
  \DOIprefix\doi{10.1063/1.1446040}.
\bibitem[{Gharib and Derango(1989)}]{Gharib1989}
\bibinfo{author}{Gharib, M.}, \bibinfo{author}{Derango, P.},
  \bibinfo{year}{1989}.
\newblock \bibinfo{title}{{A liquid film (soap film) tunnel to study
  two-dimensional laminar and turbulent shear flows}}.
\newblock \bibinfo{journal}{Physica D: Nonlinear Phenomena}
  \bibinfo{volume}{37}, \bibinfo{pages}{406--416}.
\newblock \DOIprefix\doi{10.1016/0167-2789(89)90145-0}.
\bibitem[{Govardhan and Williamson(2000)}]{Govardhan2000}
\bibinfo{author}{Govardhan, R.}, \bibinfo{author}{Williamson, C.H.},
  \bibinfo{year}{2000}.
\newblock \bibinfo{title}{{Modes of vortex formation and frequency response of
  a freely vibrating cylinder}}.
\newblock \bibinfo{journal}{Journal of Fluid Mechanics} \bibinfo{volume}{420},
  \bibinfo{pages}{85--130}.
\newblock \DOIprefix\doi{10.1017/S0022112000001233}.
\bibitem[{Griffin and Ramberg(1974)}]{Griffin1974}
\bibinfo{author}{Griffin, O.M.}, \bibinfo{author}{Ramberg, S.E.},
  \bibinfo{year}{1974}.
\newblock \bibinfo{title}{{The vortex-street wakes of vibrating cylinders}}.
\newblock \bibinfo{journal}{J. Fluid Mech.} \bibinfo{volume}{66},
  \bibinfo{pages}{553--576}.
\newblock \DOIprefix\doi{10.1017/S002211207400036X}.
\bibitem[{Griffin and Votaw(1972)}]{Griffin1972}
\bibinfo{author}{Griffin, O.M.}, \bibinfo{author}{Votaw, C.W.},
  \bibinfo{year}{1972}.
\newblock \bibinfo{title}{{The vortex street in the wake of a vibrating
  cylinder}}.
\newblock \bibinfo{journal}{Journal of Fluid Mechanics} \bibinfo{volume}{55},
  \bibinfo{pages}{31--48}.
\newblock \URLprefix
  \url{https://www.cambridge.org/core/product/identifier/S0022112072001636/type/journal{\_}article},
  \DOIprefix\doi{10.1017/S0022112072001636}.
\bibitem[{Heil et~al.(2017)Heil, Rosso, Hazel and Br{\o}ns}]{heil2017jfm}
\bibinfo{author}{Heil, M.}, \bibinfo{author}{Rosso, J.},
  \bibinfo{author}{Hazel, A.L.}, \bibinfo{author}{Br{\o}ns, M.},
  \bibinfo{year}{2017}.
\newblock \bibinfo{title}{{Topological fluid mechanics of the formation of the
  K{\'{a}}rm{\'{a}}n-vortex street}}.
\newblock \bibinfo{journal}{Journal of Fluid Mechanics} \bibinfo{volume}{812},
  \bibinfo{pages}{199--221}.
\newblock \DOIprefix\doi{10.1017/jfm.2016.792}.
\bibitem[{Heinz et~al.(2016)Heinz, S{\o}rensen, Zahle and
  Skrzypi{\'{n}}ski}]{Heinz2016}
\bibinfo{author}{Heinz, J.C.}, \bibinfo{author}{S{\o}rensen, N.N.},
  \bibinfo{author}{Zahle, F.}, \bibinfo{author}{Skrzypi{\'{n}}ski, W.},
  \bibinfo{year}{2016}.
\newblock \bibinfo{title}{{Vortex-induced vibrations on a modern wind turbine
  blade}}.
\newblock \bibinfo{journal}{Wind Energy} \bibinfo{volume}{19},
  \bibinfo{pages}{2041--2051}.
\newblock \URLprefix \url{http://doi.wiley.com/10.1002/we.1967},
  \DOIprefix\doi{10.1002/we.1967}.
\bibitem[{Huang and Larsen(2010)}]{Zhi-yong2010}
\bibinfo{author}{Huang, Z.y.}, \bibinfo{author}{Larsen, C.M.},
  \bibinfo{year}{2010}.
\newblock \bibinfo{title}{{Large Eddy Simulation on Interaction Between in-Line
  and Cross-Flow Oscillation of A Circular Cylinder}}.
\newblock \bibinfo{journal}{China Ocean Eng.} \bibinfo{volume}{24},
  \bibinfo{pages}{663--676}.
\newblock \URLprefix
  \url{http://www.chinaoceanengin.cn//article/id/fb95e7f8-9a99-41a0-abd3-9236b3e08e14}.
\bibitem[{Jia et~al.(2007)Jia, Li, Yin and Yin}]{Jia2007}
\bibinfo{author}{Jia, L.B.}, \bibinfo{author}{Li, F.}, \bibinfo{author}{Yin,
  X.Z.}, \bibinfo{author}{Yin, X.Y.}, \bibinfo{year}{2007}.
\newblock \bibinfo{title}{{Coupling modes between two flapping filaments}}.
\newblock \bibinfo{journal}{J. Fluid Mech.} \bibinfo{volume}{581},
  \bibinfo{pages}{199--220}.
\newblock \DOIprefix\doi{10.1017/S0022112007005563}.
\bibitem[{Jia and Yin(2008)}]{jia2008prl}
\bibinfo{author}{Jia, L.B.}, \bibinfo{author}{Yin, X.Z.}, \bibinfo{year}{2008}.
\newblock \bibinfo{title}{{Passive oscillations of two tandem flexible
  filaments in a flowing soap film}}.
\newblock \bibinfo{journal}{Physical Review Letters} \bibinfo{volume}{100},
  \bibinfo{pages}{228104}.
\newblock \DOIprefix\doi{10.1103/PhysRevLett.100.228104}.
\bibitem[{Jiao and Wu(2018)}]{Jiao2018}
\bibinfo{author}{Jiao, H.}, \bibinfo{author}{Wu, G.X.}, \bibinfo{year}{2018}.
\newblock \bibinfo{title}{{Free vibration predicted using forced oscillation in
  the lock-in region}}.
\newblock \bibinfo{journal}{Physics of Fluids} \bibinfo{volume}{30}.
\newblock \DOIprefix\doi{10.1063/1.5056203}.
\bibitem[{Kellay et~al.(1995)Kellay, Wu and Goldburg}]{kellay1995prl}
\bibinfo{author}{Kellay, H.}, \bibinfo{author}{Wu, X.l.},
  \bibinfo{author}{Goldburg, W.I.}, \bibinfo{year}{1995}.
\newblock \bibinfo{title}{{Experiments with turbulent soap films}}.
\newblock \bibinfo{journal}{Physical Review Letters} \bibinfo{volume}{74},
  \bibinfo{pages}{3975--3978}.
\newblock \URLprefix
  \url{https://journals.aps.org/prl/pdf/10.1103/PhysRevLett.80.277
  https://link.aps.org/doi/10.1103/PhysRevLett.74.3975},
  \DOIprefix\doi{10.1103/PhysRevLett.74.3975}.
\bibitem[{Khalak and Williamson(1999)}]{Khalak1999}
\bibinfo{author}{Khalak, A.}, \bibinfo{author}{Williamson, C.H.},
  \bibinfo{year}{1999}.
\newblock \bibinfo{title}{{Motions, Forces and Mode Transitions in
  Vortex-Induced Vibrations At Low Mass-Damping}}.
\newblock \bibinfo{journal}{Journal of Fluids and Structures}
  \bibinfo{volume}{13}, \bibinfo{pages}{813--851}.
\newblock \DOIprefix\doi{10.1006/jfls.1999.0236}.
\bibitem[{Kim and Mandre(2017)}]{Kim2017}
\bibinfo{author}{Kim, I.}, \bibinfo{author}{Mandre, S.}, \bibinfo{year}{2017}.
\newblock \bibinfo{title}{{Marangoni elasticity of flowing soap films}}.
\newblock \bibinfo{journal}{Physical Review Fluids} \bibinfo{volume}{2},
  \bibinfo{pages}{1--9}.
\newblock \DOIprefix\doi{10.1103/PhysRevFluids.2.082001},
  \href{http://arxiv.org/abs/1610.00178}{\tt arXiv:1610.00178}.
\bibitem[{Koopmann(1967)}]{Koopmann1967}
\bibinfo{author}{Koopmann, G.H.}, \bibinfo{year}{1967}.
\newblock \bibinfo{title}{{The vortex wakes of vibrating cylinders at low
  Reynolds numbers}}.
\newblock \bibinfo{journal}{J. Fluid Mech.} \bibinfo{volume}{28},
  \bibinfo{pages}{501--512}.
\newblock \DOIprefix\doi{10.1017/S0022112067002253}.
\bibitem[{Kumar and Mittal(2012)}]{Kumar2012}
\bibinfo{author}{Kumar, B.}, \bibinfo{author}{Mittal, S.},
  \bibinfo{year}{2012}.
\newblock \bibinfo{title}{{On the origin of the secondary vortex street}}.
\newblock \bibinfo{journal}{Journal of Fluid Mechanics} \bibinfo{volume}{711},
  \bibinfo{pages}{641--666}.
\newblock \DOIprefix\doi{10.1017/jfm.2012.421}.
\bibitem[{Leontini et~al.(2006)Leontini, Stewart, Thompson and
  Hourigan}]{Leontini2006}
\bibinfo{author}{Leontini, J.S.}, \bibinfo{author}{Stewart, B.E.},
  \bibinfo{author}{Thompson, M.C.}, \bibinfo{author}{Hourigan, K.},
  \bibinfo{year}{2006}.
\newblock \bibinfo{title}{{Wake state and energy transitions of an oscillating
  cylinder at low Reynolds number}}.
\newblock \bibinfo{journal}{Physics of Fluids} \bibinfo{volume}{18}.
\newblock \DOIprefix\doi{10.1063/1.2204632}.
\bibitem[{Leweke and Williamson(1998)}]{Leweke1998}
\bibinfo{author}{Leweke, T.}, \bibinfo{author}{Williamson, C.H.K.},
  \bibinfo{year}{1998}.
\newblock \bibinfo{title}{{Three-dimensional instabilities in wake
  transition}}.
\newblock \bibinfo{journal}{Eur. J. Mech. - B/Fluids} \bibinfo{volume}{17},
  \bibinfo{pages}{571--586}.
\newblock \URLprefix
  \url{https://www-sciencedirect-com.ezproxy.lib.vt.edu/science/article/pii/S0997754698800125},
  \DOIprefix\doi{10.1016/S0997-7546(98)80012-5}.
\bibitem[{Masroor et~al.(2021)Masroor,  and Stremler}]{Masroor2021}
\bibinfo{author}{Masroor, E.~Yang, W.}, , \bibinfo{author}{Stremler, M.A.},
  \bibinfo{year}{2021}.
\newblock \bibinfo{title}{Flow visualization data from experiments with an
  oscillating circular cylinder in a gravity-driven flowing soap film at low
  reynolds number}.
\newblock \bibinfo{journal}{Journal of Fluids and Structures}
  \bibinfo{note}{(in revision)}.
\bibitem[{Morse and Williamson(2009)}]{Morse2009}
\bibinfo{author}{Morse, T.L.}, \bibinfo{author}{Williamson, C.H.},
  \bibinfo{year}{2009}.
\newblock \bibinfo{title}{{Fluid forcing, wake modes, and transitions for a
  cylinder undergoing controlled oscillations}}.
\newblock \bibinfo{journal}{Journal of Fluids and Structures}
  \bibinfo{volume}{25}, \bibinfo{pages}{697--712}.
\newblock \URLprefix
  \url{http://dx.doi.org/10.1016/j.jfluidstructs.2008.12.003},
  \DOIprefix\doi{10.1016/j.jfluidstructs.2008.12.003}.
\bibitem[{Placzek et~al.(2009)Placzek, Sigrist and Hamdouni}]{Placzek2009}
\bibinfo{author}{Placzek, A.}, \bibinfo{author}{Sigrist, J.F.},
  \bibinfo{author}{Hamdouni, A.}, \bibinfo{year}{2009}.
\newblock \bibinfo{title}{{Numerical simulation of an oscillating cylinder in a
  cross-flow at low Reynolds number: Forced and free oscillations}}.
\newblock \bibinfo{journal}{Comput. Fluids} \bibinfo{volume}{38},
  \bibinfo{pages}{80--100}.
\newblock \DOIprefix\doi{10.1016/j.compfluid.2008.01.007}.
\bibitem[{Schnipper et~al.(2009)Schnipper, Andersen and Bohr}]{Schnipper2009}
\bibinfo{author}{Schnipper, T.}, \bibinfo{author}{Andersen, A.},
  \bibinfo{author}{Bohr, T.}, \bibinfo{year}{2009}.
\newblock \bibinfo{title}{{Vortex wakes of a flapping foil}}.
\newblock \bibinfo{journal}{Journal of Fluid Mechanics} \bibinfo{volume}{633},
  \bibinfo{pages}{411--423}.
\newblock \DOIprefix\doi{10.1017/S0022112009007964}.
\bibitem[{Trapeznikov(1957)}]{Trapeznikov1957}
\bibinfo{author}{Trapeznikov, A.A.}, \bibinfo{year}{1957}.
\newblock \bibinfo{title}{{Application of the method of two-dimensional
  viscosity and shear strength to the investigation of the structure and
  composition of two-sided films and surface layers in solutions of soaps and
  saponins}}.
\newblock \bibinfo{journal}{Proceedings of the 2nd International Congress on
  Surface Activity} , \bibinfo{pages}{242--258}\URLprefix
  \url{citeulike-article-id:3206466}.
\bibitem[{Vorobieff and Ecke(1999)}]{Vorobieff1999}
\bibinfo{author}{Vorobieff, P.}, \bibinfo{author}{Ecke, R.E.},
  \bibinfo{year}{1999}.
\newblock \bibinfo{title}{{Cylinder wakes in flowing soap films}}.
\newblock \bibinfo{journal}{Physical Review E - Statistical Physics, Plasmas,
  Fluids, and Related Interdisciplinary Topics} \bibinfo{volume}{60},
  \bibinfo{pages}{2953--2956}.
\newblock \DOIprefix\doi{10.1103/PhysRevE.60.2953}.
\bibitem[{Wang et~al.(2010)Wang, Tian, Jia, Lu and Yin}]{Wang2010}
\bibinfo{author}{Wang, S.Y.}, \bibinfo{author}{Tian, F.B.},
  \bibinfo{author}{Jia, L.B.}, \bibinfo{author}{Lu, X.Y.},
  \bibinfo{author}{Yin, X.Z.}, \bibinfo{year}{2010}.
\newblock \bibinfo{title}{{Secondary vortex street in the wake of two tandem
  circular cylinders at low Reynolds number}}.
\newblock \bibinfo{journal}{Phys. Rev. E} \bibinfo{volume}{81},
  \bibinfo{pages}{036305}.
\newblock \URLprefix \url{https://link.aps.org/doi/10.1103/PhysRevE.81.036305},
  \DOIprefix\doi{10.1103/PhysRevE.81.036305}.
\bibitem[{Wen and Lin(2001)}]{Wen2001}
\bibinfo{author}{Wen, C.Y.}, \bibinfo{author}{Lin, C.Y.}, \bibinfo{year}{2001}.
\newblock \bibinfo{title}{{Two-dimensional vortex shedding of a circular
  cylinder}}.
\newblock \bibinfo{journal}{Phys. Fluids} \bibinfo{volume}{13},
  \bibinfo{pages}{557--558}.
\newblock \DOIprefix\doi{10.1063/1.1338544}.
\bibitem[{Williamson and Brown(1998)}]{Williamson1998}
\bibinfo{author}{Williamson, C.H.}, \bibinfo{author}{Brown, G.L.},
  \bibinfo{year}{1998}.
\newblock \bibinfo{title}{{A series in 1/sqrt(Re) to represent the
  Strouhal-Reynolds number relationship of the cylinder wake}}.
\newblock \bibinfo{journal}{Journal of Fluids and Structures}
  \bibinfo{volume}{12}, \bibinfo{pages}{1073--1085}.
\newblock \DOIprefix\doi{10.1006/jfls.1998.0184}.
\bibitem[{Williamson and Jauvtis(2004)}]{Williamson2004a}
\bibinfo{author}{Williamson, C.H.}, \bibinfo{author}{Jauvtis, N.},
  \bibinfo{year}{2004}.
\newblock \bibinfo{title}{{A high-amplitude 2T mode of vortex-induced vibration
  for a light body in XY motion}}.
\newblock \bibinfo{journal}{European Journal of Mechanics, B/Fluids}
  \bibinfo{volume}{23}, \bibinfo{pages}{107--114}.
\newblock \DOIprefix\doi{10.1016/j.euromechflu.2003.09.008}.
\bibitem[{Williamson and Roshko(1988)}]{Williamson1988}
\bibinfo{author}{Williamson, C.H.}, \bibinfo{author}{Roshko, A.},
  \bibinfo{year}{1988}.
\newblock \bibinfo{title}{{Vortex formation in the wake of an oscillating
  cylinder}}.
\newblock \bibinfo{journal}{Journal of Fluids and Structures}
  \bibinfo{volume}{2}, \bibinfo{pages}{355--381}.
\newblock \DOIprefix\doi{10.1016/S0889-9746(88)90058-8},
  \href{http://arxiv.org/abs/arXiv:1011.1669v3}{\tt arXiv:arXiv:1011.1669v3}.
\bibitem[{Williamson(1996a)}]{Williamson1996a}
\bibinfo{author}{Williamson, C.H.K.}, \bibinfo{year}{1996}a.
\newblock \bibinfo{title}{{Three-dimensional wake transition}}.
\newblock \bibinfo{journal}{J. Fluid Mech.} \bibinfo{volume}{328},
  \bibinfo{pages}{345}.
\newblock \URLprefix
  \url{http://www.journals.cambridge.org/abstract_S0022112096008750},
  \DOIprefix\doi{10.1017/S0022112096008750}.
\bibitem[{Williamson(1996b)}]{Williamson1996}
\bibinfo{author}{Williamson, C.H.K.}, \bibinfo{year}{1996}b.
\newblock \bibinfo{title}{{Vortex Dynamics in the Cylinder Wake}}.
\newblock \bibinfo{journal}{Annu. Rev. Fluid Mech.} \bibinfo{volume}{28},
  \bibinfo{pages}{477--539}.
\newblock \URLprefix
  \url{http://www.annualreviews.org/doi/10.1146/annurev.fl.28.010196.002401},
  \DOIprefix\doi{10.1146/annurev.fl.28.010196.002401},
  \href{http://arxiv.org/abs/arXiv:1011.1669v3}{\tt arXiv:arXiv:1011.1669v3}.
\bibitem[{Williamson and Govardhan(2004)}]{Williamson2004}
\bibinfo{author}{Williamson, C.H.K.}, \bibinfo{author}{Govardhan, R.},
  \bibinfo{year}{2004}.
\newblock \bibinfo{title}{{VORTEX-INDUCED VIBRATIONS}}.
\newblock \bibinfo{journal}{Annu. Rev. Fluid Mech.} \bibinfo{volume}{36},
  \bibinfo{pages}{413--455}.
\newblock \URLprefix
  \url{http://arjournals.annualreviews.org/doi/abs/10.1146%2Fannurev.fluid.36.050802.122128},
  \DOIprefix\doi{10.1007/978-3-319-16649-0_36},
  \href{http://arxiv.org/abs/arXiv:1011.1669v3}{\tt arXiv:arXiv:1011.1669v3}.
\bibitem[{Wu et~al.(2012)Wu, Sun, Lu, Teng, Tang and Song}]{Wu2012}
\bibinfo{author}{Wu, H.}, \bibinfo{author}{Sun, D.P.}, \bibinfo{author}{Lu,
  L.}, \bibinfo{author}{Teng, B.}, \bibinfo{author}{Tang, G.Q.},
  \bibinfo{author}{Song, J.N.}, \bibinfo{year}{2012}.
\newblock \bibinfo{title}{{Experimental investigation on the suppression of
  vortex-induced vibration of long flexible riser by multiple control rods}}.
\newblock \bibinfo{journal}{J. Fluids Struct.} \bibinfo{volume}{30},
  \bibinfo{pages}{115--132}.
\newblock \DOIprefix\doi{10.1016/j.jfluidstructs.2012.02.004}.
\bibitem[{Wu et~al.(2004)Wu, Wen, Yen, Weng and Wang}]{Wu2004}
\bibinfo{author}{Wu, M.H.}, \bibinfo{author}{Wen, C.Y.}, \bibinfo{author}{Yen,
  R.H.}, \bibinfo{author}{Weng, M.C.}, \bibinfo{author}{Wang, A.B.},
  \bibinfo{year}{2004}.
\newblock \bibinfo{title}{{Experimental and numerical study of the separation
  angle for flow around a circular cylinder at low Reynolds number}}.
\newblock \bibinfo{journal}{Journal of Fluid Mechanics} \bibinfo{volume}{515},
  \bibinfo{pages}{233--260}.
\newblock \DOIprefix\doi{10.1017/S0022112004000436}.
\bibitem[{Wu et~al.(2001)Wu, Levine, Rutgers, Kellay and Goldburg}]{Wu2001}
\bibinfo{author}{Wu, X.L.}, \bibinfo{author}{Levine, R.},
  \bibinfo{author}{Rutgers, M.}, \bibinfo{author}{Kellay, H.},
  \bibinfo{author}{Goldburg, W.I.}, \bibinfo{year}{2001}.
\newblock \bibinfo{title}{{Infrared technique for measuring thickness of a
  flowing soap film}}.
\newblock \bibinfo{journal}{Rev. Sci. Instrum.} \bibinfo{volume}{72},
  \bibinfo{pages}{2467--2471}.
\newblock \DOIprefix\doi{10.1063/1.1366634}.
\bibitem[{Yang et~al.(2001)Yang, Wen and Lin}]{Yang2001}
\bibinfo{author}{Yang, T.S.}, \bibinfo{author}{Wen, C.Y.},
  \bibinfo{author}{Lin, C.Y.}, \bibinfo{year}{2001}.
\newblock \bibinfo{title}{{Interpretation of color fringes in flowing soap
  films}}.
\newblock \bibinfo{journal}{Experimental Thermal and Fluid Science}
  \bibinfo{volume}{25}, \bibinfo{pages}{141--149}.
\newblock \DOIprefix\doi{10.1016/S0894-1777(01)00087-5}.
\bibitem[{Yang and Stremler(2019)}]{Yang2019}
\bibinfo{author}{Yang, W.}, \bibinfo{author}{Stremler, M.A.},
  \bibinfo{year}{2019}.
\newblock \bibinfo{title}{{Critical spacing of stationary tandem circular
  cylinders at Re=100}}.
\newblock \bibinfo{journal}{Journal of Fluids and Structures}
  \bibinfo{volume}{89}, \bibinfo{pages}{49--60}.
\newblock \URLprefix
  \url{https://www.sciencedirect.com/science/article/pii/S088997461830817X{\#}sec4},
  \DOIprefix\doi{10.1016/J.JFLUIDSTRUCTS.2019.02.023}.
\bibitem[{Zdravkovich(1997)}]{Zdravkovich1997}
\bibinfo{author}{Zdravkovich, M.M.}, \bibinfo{year}{1997}.
\newblock \bibinfo{title}{{Flow around circular cylinders}}.
\newblock \bibinfo{publisher}{Oxford University Press}.
\bibitem[{Zhang et~al.(2017)Zhang, Dai, Abdelkefi and Wang}]{Zhang2017}
\bibinfo{author}{Zhang, L.B.}, \bibinfo{author}{Dai, H.L.},
  \bibinfo{author}{Abdelkefi, A.}, \bibinfo{author}{Wang, L.},
  \bibinfo{year}{2017}.
\newblock \bibinfo{title}{{Improving the performance of aeroelastic energy
  harvesters by an interference cylinder}}.
\newblock \bibinfo{journal}{Appl. Phys. Lett.} \bibinfo{volume}{111},
  \bibinfo{pages}{073904}.
\newblock \URLprefix \url{http://aip.scitation.org/doi/10.1063/1.4999765},
  \DOIprefix\doi{10.1063/1.4999765}.

\end{thebibliography}

\end{document}